\documentclass[journal,onecolumn]{IEEEtran}

\ifCLASSINFOpdf
  
\else

\fi
\usepackage{xcolor}
\definecolor{cool_green}{rgb}{0.0, 0.5, 0.0}

\usepackage{amsmath}
\usepackage{amssymb}
\usepackage{array}
\usepackage{tabu}
\usepackage{layout}
\usepackage{arydshln}
\usepackage{booktabs,caption}
\usepackage[flushleft]{threeparttable}
\newtheorem{theorem}{Theorem}
\newtheorem{lemma}{Lemma}
\newtheorem{definition}{Definition}
\newtheorem{fact}{Fact}

\usepackage{csquotes}
\usepackage{bbm}
\usepackage{mathtools}
\usepackage{comment}
\usepackage{physics}

\begin{document}
\title{One-shot Capacity bounds on the Simultaneous Transmission of Classical and Quantum Information }
\author{Farzin~Salek,~\IEEEmembership{Student Member,~IEEE,} Anurag~Anshu, Min-Hsiu~Hsieh,~\IEEEmembership{Senior Member,~IEEE,} Rahul~Jain, Javier~Rodr\'iguez~Fonollosa,~\IEEEmembership{Senior Member,~IEEE,}
\thanks{This work was presented in part in ISIT 2018, see \cite{FarzinISIT2018}.}
\thanks{F. Salek and J. R. Fonollosa are with the Department
of Signal Theory and Communications, Universitat Polit\`{e}cnica de Catalunya, Barcelona,
Spain, e-mail: (farzin.salek@upc.edu, javier.fonollosa@upc.edu).}
\thanks{Min-Hsiu Hsieh is with the Center for Quantum Software and Information, Sydney University of Technology, Sydney, Australia, e-mail: (min-hsiu.hsieh@uts.edu.au).}
\thanks{Anurag Anshu is with the Center for Quantum Technologies, National University of Singapore, Singapore, e-mail: (a0109169@u.nus.edu). }
\thanks{Rahul Jain is with Centre for Quantum Technologies, National University of Singapore; MajuLab, UMI 3654, Singapore and is VAJRA Adjunct Faculty, TIFR, Mumbai, India, e-mail: (rahul@comp.nus.edu.sg).}}

\maketitle

\begin{abstract}
We study the communication capabilities of a quantum channel under the most general channel model known as the \textit{one-shot} model. Unlike classical channels that can only be used to transmit classical information (bits), a quantum channel can be used for transmission of classical information, quantum information (qubits) and simultaneous transmission of classical and quantum information. In this work, we investigate the one-shot capabilities of a quantum channel for simultaneously transmitting bits and qubits. This problem was studied in the asymptotic regime for a memoryless channel where a regularized characterization of the capacity region was reported. It is known that the transmission of private classical information is closely related to the problem of quantum information transmission. We resort to this idea and find achievable and converse bounds on the simultaneous transmission of the public and private classical information. Then shifting the classical private rate to the quantum information rate leads to a rate region for simultaneous transmission of classical and quantum information. In the case of asymptotic i.i.d. setting, our one-shot result is evaluated to the known results in the literature. Our main tools used in the achievability proofs are position-based decoding and convex-split lemma.
\end{abstract} 

\begin{IEEEkeywords}
One-shot coding, channel coding, private capacity, quantum capacity
\end{IEEEkeywords}
\section{Introduction}
\IEEEPARstart{S}{hannon} modeled a noisy (classical) channel as a stochastic map $\mathcal{W}_{X\rightarrow Y}$ taking classical inputs to classical outputs according to some probability distribution, $p_{Y|X}(y|x)$ \cite{Shannon}. In his paper, he defined and computed the fundamental feature of a channel, its capacity: the amount of classical information, i.e., bits, that can be reliably transmitted from a sender to a remote receiver over a classical channel. In the limit of many independent uses of a stationary memoryless channel, Shannon showed that its capacity in bits per use of the channel is equal to the mutual information between the input and output.

Since nature is fundamentally quantum, it seemed necessary to enhance or replace Shannon's channel model with a \textit{quantum channel} model that takes quantum mechanics into account.
Many years after Shannon in the context of quantum information theory, a quantum channel is modelled by a completely-positive trace-preserving map (CPTP) with possibly different input and output Hilbert spaces. Denoted by $\mathcal{N}_{A\rightarrow B}$, a quantum channel with input and output systems $A$ and $B$ respectively, can now be used to accomplish a variety of information-processing tasks and accordingly different capacities can be defined. In the next two subsections, we review some concepts in the asymptotic and one-shot regimes.
\subsection{Memoryless and stationary channels, Asymptotic Regime}
Perhaps the most direct analogue of the capacity of a classical channel, $C(\mathcal{W})$, is the classical capacity of a quantum channel, $C(\mathcal{N})$, i.e., the highest rate (in bits per use of the channel) at which a sender can transmit classical information faithfully to a remote receiver through a quantum channel with general quantum inputs and quantum outputs. The classical capacity\footnote{hereafter, we talk about quantum channels unless otherwise specified, hence we drop the term quantum.} was independently studied in \cite{Holevo} and \cite{SW} where an achievability bound, i.e., $C(\mathcal{N})\geq \chi(\mathcal{N})$, known as HSW theorem was reported, where $\mathcal{X}(\mathcal{N})$ is the celebrated Holevo Information \cite{HolevoInfo} defined as follows:
\begin{align*}
\chi(\mathcal{N})\coloneqq\max_{p(x), \rho} I(X;B)_{\rho},
\end{align*} 
where $p(x)$ is a probability distribution, $\rho_{XB}=\sum_{x}p_{X}(x)|x\rangle\langle x|_{X}\otimes\mathcal{N}_{A\rightarrow B}(\rho_{A}^{x})$ is a bipartite quantum state and $I(X;B)_{\rho}$ is the quantum mutual information (see Definition \ref{qmi}). The classical capacity equals the regularized Holevo information, taking a limit over many copies of the channel. So unlike the classical channel, we don't fully know the capabilities of a quantum channel for transmitting classical information.  

In certain scenarios, a sender may wish to communicate classical information to a receiver by means of a quantum channel such that the information must remain secret from some third party surrounding the legitimate receiver. This information-processing task gives rise to the notion of \textit{private capacity} of a quantum channel. Cai-Winter-Yeung \cite{CWY} and Devetak \cite{Dev05} showed that the achievable rates for classical private capacity can be formulated as the difference between the Holevo information of the sender and the legitimate receiver and that of the sender and the eavesdropper(s) as given below:
\begin{align*}
\mathcal{P}(\mathcal{N})\coloneqq\max_{\rho}\left[I(X;B)_{\rho}-I(X;E)_{\rho}\right],
\end{align*}
where $\rho_{XBE}=\sum_{x}p_{X}(x)|x\rangle\langle x|_{X}\otimes\mathcal{U}_{A\rightarrow BE}^{\mathcal{N}}(\rho_{A}^{x})$ and $\mathcal{U}_{A\rightarrow BE}^{\mathcal{N}}$ is an isometric extension of the channel $\mathcal{N}_{A\rightarrow B}$. They also showed that the private capacity equals the regularized form of $\mathcal{P}(\mathcal{N})$ meaning that this ability of the quantum channel is still not fully understood.

The capacity of a quantum channel to transmit quantum information is called the quantum capacity of the channel and we represent it by $Q_{reg}(\mathcal{N})$. For a given quantum channel, one would like to understand the best rates (in terms of qubits per use of the channel) at which quantum information can be transmitted over the channel. The quantum capacity theorem was first considered in \cite{Lloyd} and later in \cite{Shorq}. Subsequently, by taking advantage of the properties of the private classical codes, Devetak \cite{Dev05} showed that the quantum capacity is given by the regularized coherent information of the channel:
\begin{align*}
Q_{reg}(\mathcal{N})\coloneqq\lim_{k\rightarrow\infty}\frac{1}{k}Q(\mathcal{N}^{\otimes k})
\end{align*}
where the coherent information is defined as $Q(\mathcal{N})\coloneqq\max_{\phi_{RA}}I(R\rangle B)_{\sigma}$ (see Definition \ref{coherentd}) and the optimization is with respect to all pure, bipartite states $\phi_{RA}$ and $\sigma_{RB}=\mathcal{N}_{A\rightarrow B}(\phi_{RA})$.

Devetak and Shor \cite{DevShor} unified the classical and quantum capacities and introduced a new information-processing task studying the simultaneously achievable rates for transmission of classical and quantum information over a quantum channel. Since we will follow the results of \cite{DevShor} closely in this paper, we mention its main theorem:\\
\begin{theorem}[\cite{DevShor}]
\label{DS}
The capacity region of $\mathcal{N}$ for simultaneous transmission of classical and quantum information is as follows:
\begin{align*}
S_{reg}(\mathcal{N})\coloneqq\lim_{k\rightarrow \infty}\frac{1}{k}S(\mathcal{N}^{\otimes k}),
\end{align*}
where $S(\mathcal{N})$ is the union, over all states $\rho_{XRB}=\sum_{x}p(x)\ketbra{x}\otimes\rho_{BR}^{x}$ arising from the channel $\mathcal{N}_{A\rightarrow B}$, i.e., for $x\in\text{supp}(p(x))$, $\rho_{RB}^{x}=\mathcal{N}(\phi_{RA}^{x})$ for pure states $\ket{\phi^{x}}_{RA}$, of the $(r, R)$ pairs obeying 
\begin{align*}
0&\leq r\leq I(X;B)_{\rho}, \\
0&\leq R\leq I(R\rangle BX)_{\rho}.
\end{align*} 
where $r$ and $R$ are the rates of the classical and quantum\footnote{It is the same for various information-processing tasks: subspace transmission, entanglement transmission or entanglement generation} information, respectively.   
\end{theorem}

The result of Devetak and Shor is generalized in \cite{MinMark2009} such that the rate of a secret key that used to achieve noiseless private capacity, enters the tradeoff. It is known that the interplay between public classical communication, private classical communication and secret key is rather analogous to how classical communication, quantum communication and entanglement interact with one another. This interaction was studied in \cite{MinMark2012} from an information-theoretic perspective and the corresponding rate regions for several realistic channels were computed.   
\subsection{General channels, One-shot Regime}
All the aforementioned capacities are originally evaluated under the assumptions that the channels are memoryless and stationary and they are available to be used arbitrarily many times. However, in many real-world scenarios, we encounter channels which are neither stationary nor memoryless. Therefore, it is of fundamental importance to think of coding schemes for the channels which fail to satisfy these assumptions. The independent channel uses are relaxed in \cite{HanVerdu93} and \cite{HayNaga03} and general channels with memory are studied in \cite{NT} and \cite{TC}, albeit these results are derived in the form of a limit such that the error probability vanishes as the number of channel uses goes to infinity. Later researchers considered \textit{single-serving} scenarios where a given channel is used only once. This approach gives rise to a high level of generality that no assumptions are made on the structure of the channel and the associated capacity is usually referred to as \textit{one-shot} capacity. 

The one-shot capacity of a classical channel was characterized in terms of min- and max-entropies in \cite{RWW06}. The one-shot classical capacity of a quantum channel is addressed by a hypothesis testing approach in \cite{MD09} and \cite{WanRnr12}, yielding expressions in terms of the generalized (R\'{e}nyi) relative entropies and a smooth relative entropy quantity, respectively. By taking advantage of two primitive information-theoretic protocols, privacy amplification and information reconciliation, authors of \cite{RnsRnr11} constructed coding schemes for one-shot transmission of public and private classical information. Their results come in terms of the min- and max-entropies. Two new tools namely position-based decoding \cite{AJW17} and convex-split lemma \cite{ADJ17}, are employed in \cite{Wil17} where one-shot achievability bounds on the public and private transmission rates are reported (note that prior to this work, one-shot bounds on the public transmission rates on both assisted and unassisted cases were reported in \cite{AJW17} and \cite{WanRnr12}, respectively). Recently, \cite{RSW17} reported tight upper and lower bounds for the one-shot capacity of the wiretap channel. This was done by proving a one-shot version of the quantum covering lemma (see \cite{AW02}) along with an operator Chernoff bound for non-square matrices. Inner and outer bounds on the one-shot quantum capacity of an arbitrary channel are studied in \cite{BusDat10}. The general scenario of \cite{BusDat10} leads to the evaluation of the quantum capacity of a channel with arbitrary correlated noise in the repeated uses of the channel. 

In this paper, we aim to study the problem of simultaneous transmission of classical and quantum information over a single use of a quantum channel. In other words, we are interested in the one-shot tradeoff between the number of bits and qubits that are simultaneously achievable.  The root of our approach is the well-known quantum capacity theorem via private classcial communication \cite{Dev05}.  The basic intuition underlying the quantum capacity is the no-cloning theorem which states that it is impossible to create an identical copy of an arbitrary unknown quantum state. We know well that associated to every quantum channel there is an environment (Eve). If Eve can learn anything about the quantum information that Alice is trying to send to Bob, Bob will not be able to retrieve this information, otherwise the no-cloning theorem would be violated. Hence, to transmit quantum information, Alice needs to store her quantum information in such subspaces of her input space that Eve does not have access to. By using this idea, Devetak \cite{Dev05} proves that a code for private classical communication can be readily translated into a code for quantum communication. Note that Devetak's proof shows the aforementioned translation in the asymptotic regime, however, one can easily check that the same holds true in the one-shot regime and the proof follows along the same lines. We provide a proof sketch in appendix B. Therefore, if we can come up with a protocol for simultaneously transmitting public and private classical information, we are able to adapt it for the simultaneous transmission of classical and quantum information.

\subsection{Techniques and Tools}
Main tools in our achievability bounds are position-based decoding and convex-split lemma. Our technique is a simple application of superposition coding in classical information theory (not to be confused with the concept of superposition in the quantum mechanics), along with convex-split lemma and position-based decoding. In this manner, we significantly differ from the technique of Devetak and Shor \cite{DevShor}, whose method was inherently asymptotic i.i.d. and could not have been adapted in the one-shot setting. 

We briefly review position-based decoding and the convex-split lemma. Assume Alice and Bob have a way of creating the following state shared between them (in other words, they have this resource at their disposal before any communication takes place):
\begin{align*}
\rho_{XA}^{\otimes |\mathcal{M}|}=\rho_{XA}^{1}\otimes...\otimes\rho_{XA}^{m}\otimes...\otimes\rho_{XA}^{|\mathcal{M}|},
\end{align*}
where Alice possesses $A$ systems and Bob has $X$ systems. Here, the positions of states is denoted by superscripts. Alice wishes to transmit the $m$-th copy of the state above through the channel $\mathcal{N}_{A\rightarrow B}$ to Bob. This induces the following state on Bob's side :
\begin{align*}
\rho_{X^{|\mathcal{M}|}B}^{m}=\rho_{X}^{1}\otimes...\otimes\rho_{XB}^{m}\otimes...\otimes\rho_{X}^{|\mathcal{M}|}.
\end{align*}
If Bob has a means by which he can distinguish between the induced states for different values of $m$ (hypotheses), which happens to be reduced to the problem of distinguishing between states $\rho_{XB}$ and $\rho_{X}\otimes\rho_{B}$, he is able to learn about the transmitted message $m$. Position-based decoding in fact, relates the communication problem to a problem in binary hypothesis testing. On the other hand, once Alice chooses the $m$-th system uniformly and sends it over the channel, the induced state on receiver side can generally be considered as: 
\begin{align*}
\frac{1}{|\mathcal{M}|}\sum_{m=1}^{|\mathcal{M}|}\rho_{X}^{1}\otimes...\otimes\rho_{XB}^{m}\otimes...\otimes\rho_{X}^{|\mathcal{M}|},
\end{align*}
 convex-split lemma argues that if the number of systems, $|\mathcal{M}|$, is almost equal to a quantity known as max-mutual information, the induced state is close to the following state 
 \begin{align*}
  \rho_{X}^{1}\otimes...\otimes\rho_{X}^{m}\otimes...\otimes\rho_{X}^{|\mathcal{M}|}\otimes\rho_{B},
 \end{align*}
  meaning that the receiver will not be able to distinguish between the induced states and the product state above, resulting in its ignorance about the chosen message $m$.

The rest of the paper is organized as follows. In Section II, we give preliminaries and definitions. A code for simultaneous transmission of public and private information is formally discussed in Section III. This section also includes our main results. Section IV is devoted to the description of the protocol as well as our achievability proof. Converse bounds are proven in section V. In Section VI, we argue how the well-known asymptotic bounds can be quickly recovered by many independent uses of a memoryless channel. We conclude the paper by a discussion in Section VII.

\section{Preliminaries}
We denote (quantum) systems by capital letters, and we use subscripts to denote the systems on which the mathematical objects are defined (we may drop the subscript if it does not lead to ambiguity). The Hilbert space corresponding to a quantum system $A$ is denoted by $\mathcal{H}_{A}$ and its dimension is shown by $|\mathcal{H}_{A}|$. Conventionally, a random variable $X$ taking on its values from some finite alphabet $\mathcal{X}$ with cardinality $|\mathcal{X}|$ can be associated with a (classical) system (which we also referred to as $X$) whose Hilbert space has orthonormal basis labeled by $x$, i.e., $\{|x\rangle\}_{x\in\mathcal{X}}$ and dimension $|\mathcal{H}_{X}|=|\mathcal{X}|$. This notation is adopted throughout the paper. The set of linear operators on $\mathcal{H}_{A}$ is denoted by $\mathcal{L}(\mathcal{H}_{A})$ and the set of non-negative operators by $\mathcal{P}(\mathcal{H}_{A})$. A state of system $A$ is a positive-semidefinite operator, i.e., $\rho_{A}\in \mathcal{P}(\mathcal{H}_{A})$, with trace equal to one. We denote the set of quantum states in $\mathcal{H}_{A}$ by $\mathcal{D}(\mathcal{H}_{A})$. The identity operator acting on $\mathcal{H}_{A}$ is shown by $\mathbbm{1}_{A}$. The trace norm of the linear operator $\rho_{A}\in\mathcal{L}(\mathcal{H}_{A})$ is defined as $\|\rho_{A}\|_{1}=\text{Tr}\{\sqrt{\rho_{A}^{\dagger}\rho_{A}}\}$ where $\rho_{A}^{\dagger}$ is the conjugate transpose of $\rho_{A}$. The support of an operator $\rho$, supp($\rho$), is defined to be the subspace orthogonal to its kernel. If the support of $\rho$ in contained in that of $\sigma$, we write $\rho\subseteq\sigma$.
Let $\mathcal{H}_{A}$ and $\mathcal{H}_{B}$ be Hilbert spaces associated to systems $A$ and $B$, respectively. We can consider the composite system of $A$ and $B$ as a single system with Hilbert space $\mathcal{H}_{A}\otimes\mathcal{H}_{B}$. For a bipartite state $\rho_{AB}\in\mathcal{D}(\mathcal{H}_{A}\otimes\mathcal{H}_{B})$, marginal systems are defined via partial trace as $\text{Tr}_{B}\{\rho_{AB}\}=\rho_{A}$ and $\text{Tr}_{A}\{\rho_{AB}\}=\rho_{B}$. For a pair of integers $i\leq j$, we define the discrete interval $[i:j]\coloneqq\{i,i+1,...,j\}$. For Hermitain operators $M$ and $N$, $M\leq N$ means that $(N-M)\in\mathcal{P}(\mathcal{H})$.

Let us consider a binary hypothesis test discriminating between the density operator $\rho_{A}$ (null hypothesis) and $\sigma_{A}$ (alternative hypothesis) where $\rho_{A}, \sigma_{A}\in \mathcal{D}(\mathcal{H}_{A})$. The task is to distinguish between the two hypotheses by means of some quantum measurement $\{T_{A}, \mathbbm{1}-T_{A}\}$ such that $0\leq T_{A}\leq \mathbbm{1}$.
The test decides in favor of $\rho_{A}$ (resp. $\sigma_{A}$) when the outcome corresponding to $T_{A}$ (resp. $\mathbbm{1}-T_{A}$) occurs. Two kinds of errors can be defined here: Type I error occurs when the true hypothesis was $\rho_{A}$ but $\sigma_{A}$ is decided and Type II error is the opposite kind of error. The error probabilities corresponding to Type I and Type II errors are respectively as follows:
\begin{align*}
\alpha(T_{A}, \rho_{A})&\coloneqq\text{Tr}\{(\mathbbm{1}-T_{A})\rho_{A}\}, \\
\beta(T_{A}, \sigma_{A})&\coloneqq\text{Tr}\{T_{A}\sigma_{A}\}.
\end{align*} 
In the setting of asymmetric hypothesis testing, the aim is to minimize $\beta(T_{A}, \sigma_{A})$ under a constraint on $\alpha(T_{A}, \rho_{A})$. This task gives rise to the definition of the hypothesis testing relative entropy defined as follows: 
\begin{definition}[Hypothesis testing relative entropy \cite{WanRnr12}, \cite{BusDat10}]
\label{htreo}
\begin{align*}
D_{H}^{\epsilon}\left(\rho_{A}\| \sigma_{A}\right)\coloneqq-\log_{2}\inf_{\substack{0\leq T_{A}\leq \mathbbm{1}, \\ \alpha(T_{A}, \rho_{A})\leq\epsilon}}\beta(T_{A}, \sigma_{A}).
\end{align*}
\end{definition}

In quantum information theory, one often needs to measure the distance between two quantum states. Let us again consider the task of distinguishing between two quantum states $\rho_{A}$ and $\sigma_{A}$ by means of a binary test operator $0\leq T_{A}\leq \mathbbm{1}$. Intuitively, the closer the states are, the harder they can be distinguished. We further assume that $\rho_{A}$ and $\sigma_{A}$ are prepared with equal probabilities. It can be easily shown that the optimal success probability in distinguishing the states equals $\frac{1}{2}(1+\max_{\substack{0\leq T\leq \mathbbm{1}}}\text{Tr}\{T_{A}(\rho_{A}-\sigma_{A})\})$. The optimization problem is evaluated as follows:
\begin{align*}
\max_{\substack{0\leq T_{A}\leq \mathbbm{1}}}\text{Tr}\{T_{A}(\rho_{A}-\sigma_{A})\}&=[\{\rho_{A}-\sigma_{A}\}_{+}(\rho_{A}-\sigma_{A})]\\ 
&\hspace*{1cm}-[\{\rho_{A}-\sigma_{A}\}_{-}(\rho_{A}-\sigma_{A})]\\
&\coloneqq\frac{1}{2}\|\rho_{A}-\sigma_{A}\|_{1},
\end{align*}
where $\{\rho_{A}-\sigma_{A}\}_{+}$ denotes the projector onto the subspace where the operator $(\rho_{A}-\sigma_{A})$ is non-negative, and $\{\rho_{A}-\sigma_{A}\}_{-}=\mathbbm{1}-\{\rho_{A}-\sigma_{A}\}_{+}$\footnote{In general, $\{\omega\}_{+}$ denotes the projector onto the positive eigenspace of $\omega$ and $\{\omega\}_{-}=\mathbbm{1}-\{\omega\}_{+}$ .}. This operational interpretation leads to a distance measure called trace distance defined below. 
\begin{definition}[Trace Distance \cite{Markbook}] 
The trace distance between two quantum states $\rho_{A}, \sigma_{A}$ is given by:
\begin{align*}
D(\rho_{A}, \sigma_{A})\coloneqq\frac{1}{2}\|\rho_{A}-\sigma_{A}\|_{1}.
\end{align*}
\end{definition}
We frequently use the following properties of the trace distance: 
\begin{itemize}
\item Trace distance is convex. For two ensembles $\{p(x), \rho_{A}^{x}\}$ and $\{p(x), \sigma_{A}^{x}\}$, where $\rho_{A}^{x}, \sigma_{A}^{x} \in \mathcal{D}(\mathcal{H}_{A})$ for all $x$, let $\rho_{XA}\coloneqq\sum_{x}p(x)\ketbra{x}\otimes\rho_{A}^{x}$ and $\sigma_{XA}\coloneqq\sum_{x}p(x)\ketbra{x}\otimes\sigma_{A}^{x}$ be the associated classical-quantum (CQ) states, respectively. Then, 
\begin{align*}
\left\|\sum_{x}p(x)\rho_{A}^{x}-\sum_{x}p(x)\sigma_{A}^{x}\right\|_{1} \leq\sum_{x}p(x)\|\rho_{A}^{x}-\sigma_{A}^{x}\|_{1}.
\end{align*}
Moreover, the following property can be easily checked:
\begin{align*}
\left\|\rho_{XA}-\sigma_{XA}\right\|_{1} =\sum_{x}p(x)\|\rho_{A}^{x}-\sigma_{A}^{x}\|_{1}.
\end{align*}
\item Trace distance is monotone non-increasing with respect to CPTP maps. That is, for quantum states $\rho$ and $\sigma$ and the map $\mathcal{N}$, the following inequality holds:
\begin{align*}
\|\mathcal{N}(\rho)-\mathcal{N}(\sigma)\|_{1}\leq\|\rho-\sigma\|_{1}.
\end{align*} 
\item Trace distance is invariant with respect to tensor-product states, meaning that for quantum states $\rho, \sigma$ and $\tau$, we have that:
\begin{align*}
\|\rho\otimes\tau-\sigma\otimes\tau\|_{1}=\|\rho-\sigma\|_{1}.
\end{align*}
\item Trace distance fulfills the triangle inequality; That is, for any three quantum states $\rho, \sigma$ and $\tau$, the following inequality holds:
\begin{align*}
 \|\rho-\sigma\|_{1}\leq\|\rho-\tau\|_{1}+\|\tau-\sigma\|_{1}.
\end{align*} 

\end{itemize}
\begin{definition}[Fidelity \cite{Uhlmann85}, \cite{Markbook}]
The fidelity between two states $\rho, \sigma\in \mathcal{D}(\mathcal{H}_{A})$ is defined as:
\begin{align*}
F(\rho, \sigma)=\|\sqrt{\rho}\sqrt{\sigma}\|_{1}.
\end{align*} 
\end{definition}
\begin{definition}[Purified Distance \cite{GLM5}, \cite{Tmchlthesis}]
Let $\rho_{A}, \sigma_{A}\in \mathcal{D}(\mathcal{H}_{A})$. The purified distance between $\rho_{A}$ and $\sigma_{A}$ is defined as:
\begin{align*}
P(\rho, \sigma)\coloneqq\sqrt{1-F(\rho, \sigma)^{2}},
\end{align*}
where $F(\rho, \sigma)$ is the fidelity. The purified distance is a metric on $\mathcal{D}(\mathcal{H})$. We use the purified distance to specify an $\epsilon$-ball around $\rho_{A}\in\mathcal{D}(\mathcal{H}_{A})$, that is $\mathcal{B}^{\epsilon}(\rho_{A})\coloneqq\{\rho^{\prime}_{A}\in \mathcal{D}(\mathcal{H}_{A}): P(\rho^{\prime}_{A}, \rho_{A})\leq\epsilon\}$.
\end{definition}
The purified distance is also monotone non-increasing with respect to quantum channels, obeys the triangle inequality and is invariant with respect to tensor product states. Moreover, the following expression shows its relation to the trace distance \cite{Tmchlthesis}:
\begin{align}
\label{rbtp}
\frac{1}{2}\|\rho-\sigma\|_{1}\leq P(\rho, \sigma)\leq\sqrt{\|\rho-\sigma\|_{1}}.
\end{align}
In addition to the hypothesis testing relative entropy, several different relative entropies and variances appear in our results and we shall consider their definitions here.
\begin{definition}[Conditional von Neumann entropy]
For a bipartite state $\rho_{AB}\in \mathcal{D}(\mathcal{H}_{A}\otimes\mathcal{H}_{B})$, we define the conditional von Neumann entropy of $A$ given $B$  as follows:
\begin{align*}
H(A|B)_{\rho}\coloneqq H(AB)_{\rho}-H(B)_{\rho},
\end{align*}
where
\begin{align*}
\qquad H(A)_{\rho}\coloneqq -\text{Tr}\{\rho_{A}\log_{2}\rho_{A}\},
\end{align*}
$H(A)_{\rho}$ is the von Neumann entropy \cite{vonNeumann}, corresponding to the Shannon entropy of the eigenvalues of $\rho_{A}$.  
\end{definition}
\begin{definition}[Quantum Mutual Information]
\label{qmi}
The quantum mutual information of a bipartite state $\rho_{AB}\in \mathcal{D}(\mathcal{H}_{A}\otimes\mathcal{H}_{B})$ is defined as follows:
\begin{align*}
I(A;B)_{\rho}\coloneqq H(A)_{\rho}+H(B)_{\rho}-H(AB)_{\rho}.
\end{align*}
The conditional quantum mutual information of a tripartite state $\rho_{ABC}\in \mathcal{D}(\mathcal{H}_{A}\otimes\mathcal{H}_{B}\otimes\mathcal{H}_{C})$ is defined in an analogous way to its classical counterpart as follows:
\begin{align*}
I(A;B|C)_{\rho}\coloneqq H(A|C)_{\rho}+H(B|C)_{\rho}-H(AB|C)_{\rho}.
\end{align*}
\end{definition}
\begin{definition}[Coherent Information]
\label{coherentd}
The coherent information of a bipartite state $\rho_{AB}\in \mathcal{D}(\mathcal{H}_{A}\otimes\mathcal{H}_{B})$ is defined as follows:
\begin{align*}
I(A\rangle B)_{\rho}\coloneqq H(B)_{\rho}-H(AB)_{\rho}.
\end{align*}
The conditional coherent information of a tripartite state $\rho_{ABC}$ is defined as $I(A\rangle B|C)_{\rho}\coloneqq H(B|C)_{\rho}-H(AB|C)_{\rho}$ and it can be shown that $I(A\rangle B|C)_{\rho}=I(A\rangle BC)_{\rho}$. In particular, for the CQ state $\rho_{XAB}=\sum_{x}p_{X}(x)|x\rangle\langle x|_{X}\otimes\rho_{AB}^{x}$, we have $ I(A\rangle BX)_{\rho}=\sum_{x}p_{X}(x)I(A\rangle B)_{\rho_{AB}^{x}}$.
\end{definition} 
\begin{definition}[Quantum Relative entropy \cite{Umegaki}]
\label{QRE}
The quantum relative entropy for $\rho_{A}, \sigma_{A}\in \mathcal{D}(\mathcal{H}_{A})$ is defined as 
\begin{align*}
D(\rho_{A}\|\sigma_{A})\coloneqq\text{Tr}\{\rho_{A}[\log_{2}\rho_{A}-\log_{2}\sigma_{A}]\},
\end{align*}
whenever $\text{supp}(\rho_{A})\subseteq\text{supp}(\sigma_{A})$ and otherwise it equals $+\infty$.
\end{definition}

\begin{fact}[Relation between the quantum relative entropy and the hypothesis testing relative entropy \cite{WanRnr12}]
\label{rbqh}
For all state $\rho_{A}$ and $\sigma_{A}$ and $\epsilon\in[0,1)$, the following inequality holds
\begin{align*}
D_{H}^{\epsilon}(\rho_{A}\|\sigma_{A})\leq\frac{1}{1-\epsilon}\left[D(\rho_{A}\|\sigma_{A})+h_{b}(\epsilon)\right],
\end{align*}
where $h_{b}(\epsilon)\coloneqq -\epsilon\log_{2}\epsilon -(1-\epsilon)\log_{2}(1-\epsilon)$ is the binary entropy function.
\end{fact}
\begin{definition}[Quantum relative entropy variance \cite{Li14}]
\label{qrev}
The quantum relative entropy variance for $\rho_{A}, \sigma_{A}\in \mathcal{D}(\mathcal{H}_{A})$ is given by:
\begin{align*}
V(\rho_{A}\|\sigma_{A})\coloneqq\text{Tr}\{\rho_{A}[\log_{2}\rho_{A}-\log_{2}\sigma_{A}-D(\rho_{A}\|\sigma_{A})]^{2}\},
\end{align*}
whenever $\text{supp}(\rho_{A})\subseteq\text{supp}(\sigma_{A})$ and $D(\rho_{A}\|\sigma_{A})$ is the quantum relative entropy.
\end{definition}
\begin{definition}[Max-relative entropy \cite{Dat09}]
\label{mre}
Max-relative entropy for $\rho_{A}, \sigma_{A}\in \mathcal{D}(\mathcal{H}_{A})$ is defined as:
\begin{align}
D_{max}(\rho_{A}\|\sigma_{A})\coloneqq\inf\left\{\lambda\in\mathbbm{R} : \rho_{A}\leq 2^{\lambda}\sigma_{A}\right\},
\end{align}
where it is well-defined if $\text{supp}(\rho_{A})\subseteq\text{supp}(\sigma_{A})$.
\end{definition}
An important property of the max-relative entropy is its monotonicity under quantum operations.
\begin{fact}[Monotonicity of max-relative entropy \cite{Dat09}]
\label{DPI}
For quantum states $\rho_{A}, \sigma_{A} \in \mathcal{D}(\mathcal{H}_{A})$ and any CPTP map $\mathcal{E}: \mathcal{L}(\mathcal{H}_{A})\rightarrow\mathcal{L}(\mathcal{H}_{B})$, it holds that 
\begin{align*}
D_{max}(\mathcal{E}(\rho_{A})\|\mathcal{E}(\sigma_{A}))\leq D_{max}(\rho_{A}\|\sigma_{A}).
\end{align*}
\end{fact}
It can be shown that the monotonicity property also holds for the hypothesis testing relative entropy in the same direction.
\begin{fact}[Relation between quantum relative entropy and max-relative entropy \cite{Dat09}]
\label{rbqm}
For quantum states $\rho_{A}, \sigma_{A} \in \mathcal{D}_{\leq}(\mathcal{H}_{A})$, it holds that 
\begin{align*}
D(\rho_{A}\|\sigma_{A})\leq D_{max}(\rho_{A}\|\sigma_{A}).
\end{align*}
\end{fact}
\begin{definition}[Smooth max-relative entropy \cite{Dat09}]
For a parameter $\epsilon\in (0, 1)$, Smooth max-relative entropy for $\rho_{A}, \sigma_{A} \in \mathcal{D}(\mathcal{H}_{A})$ is defined as:
\begin{align*}
D_{max}^{\epsilon}(\rho_{A}\|\sigma_{A})\coloneqq\inf_{\substack{\rho^{\prime}_{A}\in \mathcal{B}^{\epsilon}(\rho_{A})}}D_{max}(\rho_{A}^{\prime}\|\sigma_{A}).
\end{align*}
\end{definition}

\begin{fact}[\cite{Li14} and \cite{TH13}]
\label{factdos}
Let $\epsilon\in(0, 1)$ and $n$ be an integer. For any pair of states $\rho_{A}$ and $\sigma_{A}$ and their $n$-fold products, i.e., $\rho_{A}^{\otimes n}$ and $\sigma_{A}^{\otimes n}$, the following equations hold:
\begin{align*}
D_{H}^{\epsilon}\left(\rho_{A}^{\otimes n}\|\sigma_{A}^{\otimes n}\right)&=nD\left(\rho_{A}\|\sigma_{A}\right)\\
&+\sqrt{nV\left(\rho_{A}\|\sigma_{A}\right)}\Phi^{-1}(\epsilon)+O(\log n),
\end{align*}
\begin{align*}
D_{max}^{\epsilon}\left(\rho_{A}^{\otimes n}\|\sigma_{A}^{\otimes n}\right)&=nD\left(\rho_{A}\|\sigma_{A}\right)\\
&-\sqrt{nV\left(\rho_{A}\|\sigma_{A}\right)}\Phi^{-1}(\epsilon^{2})+O(\log n),
\end{align*}  
where $\Phi(x)=\frac{1}{\sqrt{2\pi}}\int_{-\infty}^{x}exp(-\frac{x^{2}}{2})dx$ is the cumulative distribution function for a standard Gaussian random variable and its inverse is defined as $\Phi^{-1}(\epsilon)\coloneqq\sup\{\alpha\in\mathbbm{R}|\Phi(\alpha)\leq\epsilon\}$.
\end{fact}
In the following we will define new entropic quantities that the analysis of their asymptotic behaviour requires Fact \ref{factdos} as well as a useful result in information theory known as the \textit{asymptotic equipartition property} (AEP). Let $X^{n} = (X_{1},X_{2},...,X_{n})$ be a sequence of independent and identically distributed (i.i.d.) random variables. The AEP states that for any $0 < \epsilon< 1$, any $\delta>0$ and for large enough $n$,
a randomly chosen i.i.d. sequence $x^{n}$ is with probability more than $1-\epsilon$ in a $\delta$-\textit{typical set} of sequences that satisfy
\begin{align*}
\big|\frac{1}{n}N(x_{i}|x^{n})-p(x_{i})\big|\leq\delta,
\end{align*} 
where $N(x_{i}|x^{n})$ is the number of occurrences of $x_{i}$ in the sequence $x^{n}$. To use these concepts in quantum information, the notion of \textit{typical subspace} is defined. Consider the state $\rho_{X}=\sum_{x}p(x)\ketbra{x}$. The $\delta$-typical subspace is a subspace of the full Hilbert space $\mathcal{H}_{X_{1}}\otimes...\otimes\mathcal{H}_{X_{n}}$, associated with many copies of the density operator, i.e., $\rho_{X}^{\otimes n}=\sum_{x^{n}}p(x^{n})\ketbra{x^{n}}$, that is spanned by states $\ket{x^{n}}$ whose corresponding classical sequences are $\delta$-typical. For an introduction to the quantum typicality and more on the properties of the typical subspace, we refer the reader to \cite{Markbook}.

We will present our results in terms of mutual information-like quantities defined below. We note that quantum mutual information (Definition \ref{qmi}) of a bipartite state $\rho_{AB}\in \mathcal{D}(\mathcal{H}_{A}\otimes\mathcal{H}_{B})$ can be defined alternatively by quantum relative entropy (Definition \ref{QRE}) as follows:
\begin{align*}
I(A;B)_{\rho}&\coloneqq D(\rho_{AB}\|\rho_{A}\otimes\rho_{B}).
\end{align*}
\begin{definition}[Hypothesis testing-mutual information \cite{WanRnr12}] 
\label{htma}
For a bipartite state $\rho_{AB}\in \mathcal{D}(\mathcal{H}_{A}\otimes\mathcal{H}_{B})$ and a parameter $\epsilon\in(0, 1)$, from the hypothesis testing-relative entropy (Definition \ref{htreo}), the hypothesis testing-mutual information is defined as follows:
\begin{align*}
I_{H}^{\epsilon}(A; B)_{\rho}\coloneqq D_{H}^{\epsilon}(\rho_{AB}\|\rho_{A}\otimes\rho_{B})_{\rho}.
\end{align*}
\end{definition}
\begin{definition}[Max-mutual information \cite{BCR11}] 
\label{mmi}
For a bipartite state $\rho_{AB}\in \mathcal{D}(\mathcal{H}_{A}\otimes\mathcal{H}_{B})$ and a parameter $\epsilon\in(0, 1)$, from the max-relative entropy (Definition \ref{mre}), the max-mutual information can be defined as follow:
\begin{align*}
I_{max}(A; B)_{\rho}\coloneqq D_{max}(\rho_{AB}\|\rho_{A}\otimes\rho_{B})_{\rho}.
\end{align*}
\end{definition}
\begin{definition}[Smooth max-mutual information \cite{BCR11}] 
For a bipartite state $\rho_{AB}\in \mathcal{D}(\mathcal{H}_{A}\otimes\mathcal{H}_{B})$ and a parameter $\epsilon\in(0, 1)$, from the max-mutual information (Definition \ref{mmi}), we define smooth max-mutual information as follows:
\begin{align*}
I_{max}^{\epsilon}(A; B)_{\rho}&\coloneqq\inf_{\substack{\rho^{\prime}_{AB}\in\mathcal{B}^{\epsilon}(\rho_{AB})}}D_{max}(\rho^{\prime}_{AB}\|\rho_{A}\otimes\rho_{B}).
\end{align*}
\end{definition}
The following quantity is similar to smooth max-mutual information.
\begin{definition}[smooth max-mutual information, (alternate definition) \cite{AJW17}] 
For a bipartite state $\rho_{AB}\in \mathcal{D}(\mathcal{H}_{A}\otimes\mathcal{H}_{B})$ and a parameter $\epsilon\in(0, 1)$, the smooth max-mutual information alternately can be defined as follows: 
\begin{align*}
\tilde{I}_{max}^{\epsilon}(B; A)_{\rho}&\coloneqq\inf_{\substack{\rho^{\prime}_{AB}\in\mathcal{B}^{\epsilon}(\rho_{AB})}}D_{max}(\rho^{\prime}_{AB}\|\rho_{A}\otimes\rho_{B}^{\prime}).
\end{align*}
\end{definition}
\begin{fact}[Relation between two definitions of the smooth max-mutual information, \cite{CBR14} and see lemma 2 in \cite{AJW17}]
\label{fact4}
Let $\epsilon\in(0, 1)$ and $\gamma\in(0, \epsilon)$. For a bipartite state $\rho_{AB}$, it holds that:
\begin{align*}
\tilde{I}_{max}^{\epsilon}(B; A)_{\rho}\leq I_{max}^{\epsilon-\gamma}(A; B)_{\rho}+\log_{2}\left(\frac{3}{\gamma^{2}}\right).
\end{align*}
\end{fact}
\begin{definition}[Conditional smooth hypothesis testing-mutual information]
\label{chtma}
Let $\rho_{ABX}\coloneqq\sum_{x} p_{X}(x)|x\rangle\langle x|_{X}\otimes\rho_{AB}^{x}$ be a CQ state and $\epsilon\in[0,1)$. We define
\begin{align*}
I_{H}^{\epsilon}(A;B|X)_{\rho}\coloneqq \max_{\substack{\rho'}} \min_{\substack{x\in\text{supp}\left(\rho'_{X}\right)}} I_{H}^{\epsilon}(A;B)_{\rho_{AB}^{x}},
\end{align*}
where maximization is over all $\rho'_{X}=\sum_{x}p'_{X}(x)|x\rangle\langle x|_{X}$ satisfying $P(\rho'_{X}, \rho_{X})\leq\epsilon$.
\end{definition}
\begin{definition}[Conditional smooth max-mutual information\footnote{Conditional alternate smooth max-information can be defined in the same way.}]
\label{casmma}
Let $\rho_{ABX}\coloneqq\sum_{x} p_{X}(x)|x\rangle\langle x|_{X}\otimes\rho_{AB}^{x}$ be a CQ state and $\epsilon\in[0,1)$. The conditional smooth max-mutual information is defined as follows:
\begin{align*}
I_{max}^{\epsilon}(A;B|X)_{\rho}\coloneqq \min_{\substack{\rho'}} \max_{\substack{x\in\text{supp}\left(\rho'_{X}\right)}} I_{max}^{\epsilon}(A;B)_{\rho_{AB}^{x}},
\end{align*}
where minimization is over all $\rho'_{X}=\sum_{x}p'_{X}(x)|x\rangle\langle x|_{X}$ satisfying $P(\rho'_{X}, \rho_{X})\leq\epsilon$.
\end{definition}
\begin{lemma}
\label{l1}
Let $\rho_{XAB}=\sum_{x}p(x)|x\rangle\langle x|_{X}\otimes\rho_{AB}^{x}$. Then the following holds:
\begin{align*}
\lim_{n\rightarrow\infty}\frac{1}{n} I_{H}^{\epsilon}(A^{\otimes n};B^{\otimes n}|X^{n})_{\rho^{\otimes n}}=I(A;B|X)_{\rho},
\end{align*}
\begin{IEEEproof}
The following is easily seen from the definition,
\begin{align*}
&I_{H}^{\epsilon}(A^{\otimes n};B^{\otimes n}|X^{n})_{\rho^{\otimes n}}\\
&\hspace*{1cm}\coloneqq\max_{\rho_{X^{n}}^{\prime}}\min_{x^{n}\in supp(\rho_{X^{n}}^{\prime})}D_{H}^{\epsilon}(\rho_{A^{n}B^{n}}^{x^{n}}\|\rho_{A^{n}}^{x^{n}}\otimes\rho_{B^{n}}^{x^{n}}).
\end{align*}
In order to be able to apply the asymptotic results given in Fact \ref{factdos}, we first produce $\rho_{X^{n}}^{\prime}$ by projecting $\rho_{X}^{ \otimes n}$ onto its typical subspace and properly normalize it. We know that the resulting state is close to the initial product state. Conditioned on a particular typical sequence $x^{n}$, the state $\rho_{A^{n}B^{n}}^{x^{n}}$ is in fact a tensor-product state that can be written as $\rho_{AB}^{x(1)}\otimes...\otimes\rho_{AB}^{x(i)}\otimes...\otimes\rho_{AB}^{x(n)}$ in which $x(i)$, $i\in [1:n]$ indicates the $i$-th index in the sequence $x^{n}$. From the definition of the typical sequences, we know that for $n$ large enough, each realization $x$ appears almost $np(x)$ times in each sequence. Hence, for any $\delta\geq 0$, as $n\rightarrow\infty$, by using Fact \ref{factdos} for each chosen sequence, the multi-letter formula above can be written as shown by (\ref{asymptotichp}) where $x_{i}$, $i\in[1:|\mathcal{X}|]$ denotes an element of the alphabet $\mathcal{X}$ and the second equality follows from Fact \ref{factdos} and the fully quantum AEP \cite{TH13}.
\begin{figure*}[!t]
\begin{align}
\nonumber
&\lim_{n\rightarrow\infty}\frac{1}{n}D_{H}^{\epsilon}(\rho_{AB}^{x^{n}}\|\rho_{A}^{x^{n}}\otimes\rho_{B}^{x^{n}})=\lim_{n\rightarrow\infty}\frac{1}{n}D_{H}^{\epsilon}\left(\rho_{AB}^{np(x_{1})\pm\delta}\otimes...\otimes\rho_{AB}^{np(x_{|\mathcal{X}|})\pm\delta}\|(\rho_{A}^{x_{1}}\otimes\rho_{B}^{x_{1}})^{\otimes np(x_{1})\pm\delta}\otimes...\otimes(\rho_{A}^{x_{|\mathcal{X}|}}\otimes\rho_{B}^{x_{|\mathcal{X}|}})^{\otimes np(x_{|\mathcal{X}|})\pm\delta}\right)\\
&=\lim_{n\rightarrow\infty}\frac{1}{n}\sum_{i=1}^{|\mathcal{X}|}\big(np(x_{i})\pm\delta\big)D(\rho_{AB}^{x_{i}}\|\rho_{A}^{x_{i}}\otimes\rho_{B}^{x_{i}})=\sum_{x=1}^{|\mathcal{X}|}p(x)D(\rho_{AB}^{x}\|\rho_{A}^{x}\otimes\rho_{B}^{x})\coloneqq I(A;B|X)_{\rho}.
\label{asymptotichp}
\end{align}
\hrulefill
\end{figure*}
\end{IEEEproof}
\end{lemma}
\begin{lemma}
\label{l2}
Let $\rho_{XAB}=\sum_{x}p(x)|x\rangle\langle x|_{X}\otimes\rho_{AB}^{x}$. Then the following holds.
\begin{align*}
\lim_{n\rightarrow\infty}\frac{1}{n} I_{max}^{\epsilon}(A^{\otimes n};B^{\otimes n}|X^{n})_{\rho^{\otimes n}}=I(A;B|X)_{\rho}
\end{align*}
\begin{IEEEproof}
The proof is very similar to that of Lemma \ref{l1}. It employs the properties of the typical sequences as well as the fully quantum asymptotic equipartition property (AEP) for smooth max-mutual information \cite{TH13}.
\end{IEEEproof}
\end{lemma}
\begin{lemma}
\label{lh}
For a CQ state $\rho_{XAB}=\sum_{x}p(x)|x\rangle\langle x|_{X}\otimes\rho_{AB}^{x}$, the following inequality is true.
\begin{align*}
I_{H}^{\epsilon}(A;B|X)_{\rho}\leq\frac{1}{1-\epsilon}\left(I(A;B|X)+h_{b}(\epsilon)\right)_{\rho}.
\end{align*}
\begin{IEEEproof}
Considering the definition of the conditional hypothesis testing-mutual information and the fact that
\begin{align*}
\min_{x}D_{H}^{\epsilon}(\rho_{AB}^{x}\|\rho_{A}^{x}\otimes\rho_{B}^{x})\leq\sum_{x}p(x)D_{H}^{\epsilon}(\rho_{AB}^{x}\|\rho_{A}^{x}\otimes\rho_{B}^{x}),
\end{align*}
and also from Fact \ref{rbqh} for all $x$, we have:
\begin{align*}
D_{H}^{\epsilon}(\rho_{AB}^{x}\|\rho_{A}^{x}\otimes\rho_{B}^{x})\leq\frac{1}{1-\epsilon}\left(D(\rho_{AB}^{x}\|\rho_{A}^{x}\otimes\rho_{B}^{x})+h_{b}(\epsilon)\right),
\end{align*}
by plugging into the the aforementioned inequality, we can get the result. We note than in order for the above to be true, we should have $\rho_{X}^{\prime}\subseteq\rho_{X}$. However, in case $\rho_{X}^{\prime}$ goes beyond the support of $\rho_{X}$, it can be projected onto the support of $\rho_{X}$. Since $P(\rho_{X}^{\prime},\rho_{X})\leq\epsilon$, from the monotonicity of the purified distance, it can be seen that the state after being projected will remain $\epsilon$-close to the initial state. 
\end{IEEEproof}
\end{lemma}
\begin{lemma}
\label{lm}
Let $\rho_{XAB}=\sum_{x}p(x)|x\rangle\langle x|_{X}\otimes\rho_{AB}^{x}$. The following inequality holds.
\begin{align*}
&I_{max}^{\epsilon}(A;B|X)_{\rho}\geq I(A;B|X)_{\rho}\\
&\hspace*{3cm}-2\epsilon\log|\mathcal{H}_{A}|-2(1+\epsilon)h_{b}(\frac{\epsilon}{1+\epsilon}).
\end{align*}
\begin{IEEEproof}
In the the following simple inequality: 
\begin{align}
\label{pd1}
\max_{x}D_{max}^{\epsilon}(\rho_{AB}^{x}\|\rho_{A}^{x}\otimes\rho_{B}^{x})\geq\sum_{x}p(x)D_{max}^{\epsilon}(\rho_{AB}^{x}\|\rho_{A}^{x}\otimes\rho_{B}^{x}),
\end{align}
we have to deal with $D_{max}^{\epsilon}(\rho_{AB}^{x}\|\rho_{A}^{x}\otimes\rho_{B}^{x})$ and try to bound it from below. Let $\bar{\rho}_{AB}^{x}$ be the state achieving the minimum in the definition of $D_{max}^{\epsilon}(\rho_{AB}^{x}\|\rho_{A}^{x}\otimes\rho_{B}^{x})$, hence
\begin{align*}
D_{max}^{\epsilon}(\rho_{AB}^{x}\|\rho_{A}^{x}\otimes\rho_{B}^{x})\geq D_{max}(\bar{\rho}_{AB}^{x}\|\bar{\rho}_{A}^{x}\otimes\bar{\rho}_{B}^{x})
\end{align*}
where $P(\rho_{AB}^{x},\bar{\rho}_{AB}^{x})\leq\epsilon$.
From Fact \ref{rbqm} we further know that $D_{max}(\bar{\rho}_{AB}^{x}\|\bar{\rho}_{A}^{x}\otimes\bar{\rho}_{B}^{x})\geq D(\bar{\rho}_{AB}^{x}\|\bar{\rho}_{A}^{x}\otimes\bar{\rho}_{B}^{x})$. Now we deploy Alicki-Fannes-Winter (AFW) inequality \cite{AFW1} (an improvement over \cite{AFW}) for the quantum mutual information saying that: (from the relation between the purified and trace distances, we know that $\frac{1}{2}\|\rho_{AB}^{x}-\bar{\rho}_{AB}^{x}\|\leq\epsilon$)
\begin{align*}
&D(\bar{\rho}_{AB}^{x}\|\bar{\rho}_{A}^{x}\otimes\bar{\rho}_{B}^{x})\geq D(\rho_{AB}^{x}\|\rho_{A}^{x}\otimes\rho_{B}^{x})\\
&\hspace*{3.5cm}-2\epsilon\log|\mathcal{H}_{A}|-2(1+\epsilon)h_{2}(\frac{\epsilon}{1+\epsilon}).
\end{align*}
Therefore, 
\begin{align*}
&D_{max}^{\epsilon}(\rho_{AB}^{x}\|\rho_{A}^{x}\otimes\rho_{B}^{x})\geq D(\rho_{AB}^{x}\|\rho_{A}^{x}\otimes\rho_{B}^{x})\\
&\hspace*{3.5cm}-2\epsilon\log|\mathcal{H}_{A}|-2(1+\epsilon)h_{2}(\frac{\epsilon}{1+\epsilon}),
\end{align*}
and plugging back into the right-hand side of (\ref{pd1}), we well get the desired result.
\end{IEEEproof}
\end{lemma}
 
\begin{lemma}[Convex-split lemma \cite{ADJ17}]
Fix $\epsilon\in (0, 1)$ and $\delta\in (0, \epsilon)$. Let $\rho_{AB}\in\mathcal{D}(\mathcal{H}_{A}\otimes\mathcal{H}_{B})$ and define the state $\tau_{A_{1}...A_{K}B}$ as follows:
\begin{align*}
\tau&_{A_{1}...A_{|\mathcal{K}|}B}\\
&=\frac{1}{|\mathcal{K}|}\sum_{k=1}^{|\mathcal{K}|}\rho_{A_{1}}\otimes...\otimes\rho_{A_{k-1}}\otimes\rho_{A_{k}B}\otimes\rho_{A_{k+1}}\otimes...\otimes\rho_{A_{|\mathcal{K}|}}.
\end{align*}
If
\begin{align*}
\log_{2}|\mathcal{K}|\geq\tilde{I}_{max}^{\sqrt{\epsilon}-\delta}(B; A)_{\rho}+2\log_{2}\left(\frac{1}{\delta}\right),
\end{align*}
then
\begin{align*}
P(\tau_{A_{1}...A_{|\mathcal{K}|}B}, \rho_{A_{1}}\otimes...\otimes\rho_{A_{k}}\otimes...\otimes\rho_{A_{|\mathcal{K}|}}\otimes\tilde{\rho}_{B})\leq\sqrt{\epsilon},
\end{align*}
where $\tilde{\rho}_{B}$ is the marginal of some state $\tilde{\rho}_{AB} \in\mathcal{B}^{\sqrt{\epsilon}-\delta}(\rho_{AB})$. The above smooth version of convex-split lemma is taken from \cite{Wil17}, which improved the error parameters in the smooth version given in \cite{AJW17}.
\end{lemma}
\begin{lemma}[Hayashi-Nagaoka operator inequality \cite{HayNaga03}]
\label{HN}
Let $T, S\in\mathcal{P}(\mathcal{H}_{A})$ such that $(\mathbbm{1}-S)\in\mathcal{P}(\mathcal{H}_{A})$. Then for all constants $c>0$, the following inequality holds:
\begin{align*}
\mathbbm{1}-(S+T)^{-\frac{1}{2}}S&(S+T)^{-\frac{1}{2}}\\
&\leq(1+c)(\mathbbm{1}-S)+(2+c+c^{-1})T.
\end{align*}
\end{lemma}
\begin{lemma}[Gentle measurement lemma \cite{W99a}]
Let $\rho_{A}\in\mathcal{D}(\mathcal{H}_{A})$ and $0\leq\Lambda_{A}\leq\mathbbm{1}$ be a measurement operator. If the measurement operator decides in favor of $\rho_{A}$ with high probability, $\text{Tr}\{\Lambda_{A}\rho_{A}\}\geq1-\epsilon$ for $\epsilon\in[0,1]$, then
\begin{align*}
\left\|\rho_{A}-\sqrt{\Lambda_{A}}\rho_{A}\sqrt{\Lambda_{A}}\right\|_{1}\leq 2\sqrt{\epsilon}.
\end{align*} 
\end{lemma}

We note that here we consider quantum communication channels with quantum input and outputs. One may consider channels with classical inputs and quantum outputs, i.e., CQ channels. In this case, an encoder has to be prepended to the CQ channel such that it associates a particular input state to every classical input.

\section{Problem Statement And Main Results}
In this section, we first define a simultaneous public-private one-shot code, then we present our main results. Latter, we discuss the translation of the  public-private code to a classical-quantum code. 
Two classical messages $(m, \ell)\in\mathcal{M}\times\mathcal{L}$ are to be transmitted from a sender to a receiver in the presence of an eavesdropper by using a quantum channel only once, i.e., one-shot communication is considered. The sender Alice, wishes to reliably communicate a public message $m$ and (simultaneously)  a private message $\ell$ to the legitimate receiver Bob such that $\ell$ must not be leaked to the eavesdropper Eve. The quantum (wiretap) channel to be used by three parties is denoted by $\mathcal{N}_{A\rightarrow BE}$ and it takes quantum states from $\mathcal{H}_{A}$ to $\mathcal{H}_{B}\otimes\mathcal{H}_{E}$
where Alice is assumed to control the input system $A$ and systems $B$ and $E$ are outputs received by Bob and Eve, respectively. Let $M$ and $L$ be the random variables\footnote{$M$ and $L$ basically are registers which hold the public and private messages, respectively. Here with slightly abuse of notation, we refer to them as random variables to which, corresponding classical states can be tied.} corresponding to Alice's choices of the public and private messages, respectively\footnote{In the literature, for example \cite{Csiszar-Korner}, the public and private messages are referred to as the common and confidential messages, respectively. If Eve were to receive the common message, it could have been considered without jeopardizing the confidential message. Indeed, as we will see, the secrecy analysis is guaranteed assuming Eve has detected the common (or the public) message.}. We formally define a one-shot simultaneous public-private code in the following.
\begin{definition}
\label{maind}
Fix $\epsilon$, $\epsilon' \in (0,1)$ and let $r$ and $R$ be the rates of the public and private messages, respectively (i.e., $|\mathcal{M}|=2^{r}$ and $|\mathcal{L}|=2^{R}$). A one-shot $(r, R, \epsilon, \epsilon')$- simultaneous public-private code for the channel $\mathcal{N}_{A\rightarrow BE}$ consists of 
\begin{itemize}
\item  An encoding operation by Alice $\mathcal{E}:ML\rightarrow \mathcal{D}(\mathcal{H}_{A})$ such that
\begin{align}
\label{d2}
\forall m\in \mathcal{M},\qquad\frac{1}{2}\|\rho_{LE}^{m} - \rho_L \otimes \tilde{\rho}_E^{m}\|_{1} \leq \epsilon',
\end{align}
where for each message $m$, $\rho_{LE}^{m}$ and $\rho_L$ are appropriate marginals of the state $\rho_{LBE}^{m}=\frac{1}{|\mathcal{L}|}\sum_{\ell=1}^{|\mathcal{L}|}|\ell\rangle\langle\ell|\otimes\mathcal{N}(\mathcal{E}(m,\ell))$ and $\tilde{\rho}_E^{m}$ can be any arbitrary state.
\item A decoding operation by Bob $\mathcal{D}:\mathcal{D}(\mathcal{H}_{B})\rightarrow \hat{M}\hat{L}$ such that
\begin{align}
\label{d1}
Pr\left((\hat{M},\hat{L})\neq(M,L)\right)\leq \epsilon,
\end{align}
where $\hat{M}$ and $\hat{L}$ denote the estimates of the public and private messages, respectively.
\end{itemize}
\end{definition}
A rate pair $(r, R)$ is said to be $(\epsilon,\epsilon')$-achievable if there exist encoding and decoding maps $(\mathcal{E},\mathcal{D})$ such that (\ref{d2}) and (\ref{d1}) are fulfilled. For a given $(\epsilon,\epsilon')$, the one-shot capacity region for the simultaneous transmission of public and private information of the channel $\mathcal{N}$, $\mathcal{C}^{\epsilon,\epsilon'}(\mathcal{N})$, is the closure of all achievable rate pairs in a $(r, R, \epsilon, \epsilon')$ coding scheme. In this work, our aim is to find upper and lower bounds on $\mathcal{C}^{\epsilon,\epsilon'}(\mathcal{N})$.

In the following, we first have Theorem \ref{achievability tm} that establishes a lower bound on $\mathcal{C}^{\epsilon,\epsilon'}(\mathcal{N})$ referred to as achievability and then Theorem \ref{converse} that states an upper bound on $\mathcal{C}^{\epsilon,\epsilon'}(\mathcal{N})$, i.e., the converse. This section ends with a discussion about the translation of the private classical capacity to the quantum capacity in one-shot regime. 
\begin{theorem}[Achievability]
\label{achievability tm}
For any fixed $\epsilon\in(0,1), \epsilon'\in(0,1),$ and $\delta, \delta'$ such that $\delta\in (0, \epsilon), \delta'\in (0, \epsilon '),$ there exists a one-shot $(r,R,3\epsilon+2\sqrt{\epsilon}+\sqrt{\epsilon'},2(\epsilon+\sqrt{\epsilon})+\sqrt{\epsilon'})$ code for the channel $\mathcal{N}_{A\rightarrow BE}$ if  the twin $(r,R)$ satisfies the following bounds:
\begin{align*}
r&\leq I_{H}^{\epsilon-\delta}\left(X;B\right)_{\rho}-\log_{2}(\frac{4\epsilon}{\delta^{2}}),\\
R&\leq I_{H}^{\epsilon-\delta}\left(Y;B|X\right)_{\rho}-\tilde{I}_{max}^{\sqrt{\epsilon'}-\delta'}(Y;E|X)_{\rho}\\
&\hspace*{4cm}-\log_{2}(\frac{4\epsilon}{\delta^{2}})-2\log_{2}(\frac{1}{\delta'}),
\end{align*}
for some quantum state $\rho$ arising from the channel. We call the region above $\mathcal{C}_{a}(\mathcal{N})$, therefore, we have 
\begin{align*}
\mathcal{C}_{a}(\mathcal{N})\subseteq\mathcal{C}^{3\epsilon+2\sqrt{\epsilon}+\sqrt{\epsilon'},2(\epsilon+\sqrt{\epsilon})+\sqrt{\epsilon'}}(\mathcal{N}).
\end{align*}
\end{theorem}

\begin{theorem}[Converse]
\label{converse}
For any fixed $\epsilon\in(0,1), \epsilon'\in(0,1)$, every one-shot $(r,R,\epsilon,\epsilon')$ public-private code for the channel $\mathcal{N}_{A\rightarrow BE}$, must satisfy the following inequalities:
\begin{align*}
r\leq& I_{H}^{\epsilon}(X;B)_{\rho},\\
R\leq& I_{H}^{\sqrt{\epsilon}}(Y;B|X)_{\rho}-I_{max}^{\sqrt{2\epsilon'}}(Y;E|X)_{\rho},
\end{align*}  
for some state $\rho_{XYBE}=\sum_{x,y}p(x,y)|x\rangle\langle x|\otimes |y\rangle\langle y|\otimes \rho_{BE}^{x,y}$. We refer to this region as $\mathcal{C}_{c}(\mathcal{N})$. In fact, we have $\mathcal{C}^{\epsilon,\epsilon'}(\mathcal{N})\subseteq\mathcal{C}_{c}(\mathcal{N})$.
\end{theorem}

Once there is a code for simultaneous transmission of public and private classical information, this code can be translated into a coherent code that is capable of transmitting classical and quantum information simultaneously. In other words, the rate pair (public classical, private classical) can be shifted to the rate pair (public classical, quantum) (or simply (classical, quantum)). We can then translate our one-shot (public, private) code to a one-shot (classcial, quantum) code. Note that the proof is implicit in findings of Devetak \cite{Dev05} such that one can mimic his procedure to see the result in one-shot setting. Henceforth, we have a one-shot code for simultaneous transmission of classical and quantum information. 

By evaluating the asymptotic behaviour of the rate region given by Theorem \ref{achievability tm} and Theorem \ref{converse} (Section VI), we recover Theorem 1 of  \cite{DevShor}, the well-known result of Devetak and Shor, as a corollary.

\section{Achievability}

We consider a general quantum channel which is prepended by an encoder (modulator) that associates a particular input state to every classical input pair. In this sense, Alice can be thought of as being in possession of an ensemble $\{p_{X,Y}(x,y),\omega_{A}^{x,y}\}$ such that the input distribution $p(x,y)$ and the encoder need to be optimized over to get our capacity results. In our protocol, Bob runs two successive decodings, his first decoder has $|\mathcal{M}|$ possible classical outputs as well as a post-measurement quantum state. His second decoder takes the resulted states of the first decoder and its output is a classical system of dimension $|\mathcal{L}|$. Before we get into achievability proof, we describe our protocol. 
\subsection{Protocol description}
Fix a joint probability distribution $p_{X,Y}(x,y)$ over the finite alphabets $\{\mathcal{X}\times\mathcal{Y}\}$, $\epsilon,\epsilon'\in(0,1)$, $\delta\in(0,\epsilon)$, $\delta'\in(0,\sqrt{\epsilon'})$ and $\rho_{XYBE}=\sum_{x,y}p(x,y)|x\rangle\langle x|\otimes |y\rangle\langle y|\otimes \rho_{BE}^{x,y}$. Let
\begin{align*}
&r\leq I_{H}^{\epsilon-\delta}\left(X;B\right)_{\rho}-\log_{2}(\frac{4\epsilon}{\delta^{2}}), \\
&R+\tilde{R}\leq I_{H}^{\epsilon-\delta}\left(Y;B|X\right)_{\rho}-\log_{2}(\frac{4\epsilon}{\delta^{2}}), \\
&\tilde{R}\geq\tilde{I}_{max}^{\sqrt{\epsilon'}-\delta'}(E;Y|X)_{\rho}+2\log_{2}(\frac{1}{\delta'}).
\end{align*}
We choose $|\mathcal{M}|=2^{r}$, $|\mathcal{L}|=2^{R}$ and $|\mathcal{K}|=2^{\tilde{R}}$ implying that $r$ and $R$ denote our public and private rates, respectively and $|\mathcal{K}|$ stands for the size of a local key, a uniformly distributed random variable $K$, used by Alice for obfuscation purpose. Let the sender Alice, legitimate receiver Bob and Eve be connected by means of a quantum (wiretap) channel $\mathcal{N}^{A\rightarrow BE}$.
 
 Alice wants to convey to Bob, in a single use of a quantum channel, a classical message $m\in \mathcal{M}$ and simultaneously, a private classical message $\ell \in \mathcal{L}$ where both messages are uniformly distributed on their corresponding sets. The message $m$ is public, meaning that Bob has to be able to decode it correctly with small probability of error. On the other hand, message $\ell$ is private and while Bob has to receive it with negligible error probability, it must be kept secret from Eve. We clarify that our definition of public and private messages is the same as in \cite{DevShor} and these correspond respectively to common and confidential messages defined in \cite{Csiszar-Korner}. The position-based decoding is employed in order to accomplish this information-processing task, therefore before communication begins, Alice, Bob and Eve share the state given in (\ref{Srd-state}), where Alice controls the system $A$, Bob has systems $(X,Y)$ and Eve is in possession of $(X',Y')$ systems. Our coding scheme is, in spirit, inferred from the well-known superposition coding in classical information theory \cite{Coverbook}. We can think of the state (\ref{Srd-state}) as the superposition of two states, each of which is use to accomplish a certain part of the task. There are $|\mathcal{M}|$ bins in the first place, inside each of them, there are $|\mathcal{L}||\mathcal{K}|$ states that are divided into $|\mathcal{L}|$ bins, again inside each one  there are $|\mathcal{K}|$ states.
 \begin{figure*}[!t]
\begin{align}
\rho_{XX'(AYY')^{\otimes|\mathcal{L}||\mathcal{K}|}}^{\otimes |\mathcal{M}|}\coloneqq\left(\sum_{x}p(x)|x\rangle\langle x|_{X}\otimes |x\rangle\langle x|_{X'}
\otimes \left(\sum_{y}p(y|x)|y\rangle\langle y|_{Y} \otimes |y\rangle\langle y|_{Y'}\otimes\rho_{A}^{x,y}\right)^{\otimes |\mathcal{L}||\mathcal{K}|}\right)^{\otimes |\mathcal{M}|}.
\label{Srd-state}
\end{align}
\hrulefill
\end{figure*}
 
Upon receiving the message pair $(m, \ell)$, Alice goes to the $m$-th copy of $\rho_{XX'(AYY')^{\otimes|\mathcal{L}||\mathcal{K}|}}^{\otimes |\mathcal{M}|}$. There she runs the protocol for the private capacity, by considering $|\mathcal{L}||\mathcal{K}|$ copies and choosing a system $A$ uniformly at random from the $\ell$-th bin. 
Upon receiving $B$, Bob performs a position-based decoding to obtain the public message $m$ (and hence the correct copy of $\rho_{XX'(AYY')^{\otimes|\mathcal{L}||\mathcal{K}|}}$). The choice of the rate for public message $r$ ensures that this is possible and gentle measurement lemma ensures that the quantum state of the correct copy of $\rho_{XX'(AYY')^{\otimes|\mathcal{L}||\mathcal{K}|}}$ is almost unchanged after Bob's decoding.

To decode $\ell$, Bob performs another position-based decoding conditioned on $X$, meaning that having found the correct copy of $\rho_{XX'(AYY')^{\otimes|\mathcal{L}||\mathcal{K}|}}$ used in the transmission, Bob applies  a decoder that depends on $X$, and it works for all $x\in\mathcal{X}$. For this strategy, Bob first appeals to the definition of the conditional smooth hypothesis testing-mutual information, to assume that the distribution over X was $p'(x)$ (achieving the infimum in the definition) with negligible error. Then for $x\in\text{supp}(\rho_{X'})$, he performs position-based decoding. The choice of $R+\tilde{R}$ guarantees the successful decoding for every $x$ and at the same time, the security criterion is ensured from the fact that even if Eve is aware of the correct copy of $\rho_{XX'(AYY')^{\otimes|\mathcal{L}||\mathcal{K}|}}$, the condition that convex-split lemma imposes on $|\mathcal{K}|$, gives her very small information about $\ell$ for every $x\in\text{supp}(\rho_{X'})$ (where here $\rho_{X'}=\sum_{x}p_{X'}(x)|x\rangle\langle x|_{X'}$ and $p_{X'}(x)$ is the distribution achieving the infimum in the alternate definition of conditional smooth max-mutual information). Now we can derandomize the protocol by fixing the values in corresponding systems. Upon derandomization, the code is publicly available. 

Before we proceed to the error analysis of the direct part, we make the following remarks. The state that is fed into the second decoder differs from the original state although negligibly, this adds to the error probability of the private message. Moreover, since successive cancellation decoding is being performed, in the event of a failure of the first decoder, the second decoder will fail as well. We also take the contribution of this event into account. Moreover, note that there is just one decoding map in general, Bob's (two) separate decodings are just a property of our protocol. 

 \subsection{Achievability Proof}
As is learned in the preceding subsection, we start with a randomness assisted protocol and derandomize it later. We get started on our proof by introducing the encoder and the decoders. We then analyze the average error probability of the public message. Likewise, we inspect the second decoder and analyze the average error probability of the private message. Finally, we study the secrecy requirement. 

In the achievability part of our randomness assisted code, for the private message, we stick to a single criterion known as \textit{privacy error} introduced in \cite{KMPJ2009}, \cite{MMM2017} and \cite{Wil17}. The general idea is to merge the secrecy of the private message (\ref{d2}) as well as its error probability (\ref{d1}) into one single criterion. While this idea was used in \cite{KMPJ2009} and \cite{MMM2017} in understanding upper bounds for private communication protocols, it had not been used in an achievability proof prior to \cite{Wil17}. We should note that the main advantage of dealing with single criterion reveals when the code is to be derandomized. Our procedure is that we analyze the error probability of Bob in detecting the private message separately from keeping Eve ignorant. This leads to two separate criteria and then the separate criteria are merged into one single criterion. It is clear that if the joined criterion is satisfied, each of the single criteria is also fulfilled. After we prove the correctness of these criteria for the randomness assisted code, we immediately proceed to derandomize the code in the succeeding step that the unassisted criteria set out by Definition \ref{maind} can be inferred. The derandomization involves some procedures that appeared in \cite{Wil17} and \cite{HQM2017}.

Alice, Bob and Eve are allowed to share some quantum state among themselves. Moreover, Alice has access to a source of uniform dummy randomness given in random variable $K$. Further, let $\tilde{R}=\log_{2}|\mathcal{K}|$. The state initially shared between three parties is given by equation (\ref{Srd-state}),
where Alice possesses the quantum systems $A$, Bob possesses the classical systems $(X, Y)$ and Eve has the classical systems $(X',Y')$. For ease of notation, we further define $\Upsilon_{T_{X}T_{X'}T_{A}T_{Y}T_{Y'}}\coloneqq\rho_{XX'(AYY')^{\otimes|\mathcal{L}||\mathcal{K}|}}^{\otimes |\mathcal{M}|}$ with it being clear that for example $\Upsilon_{T_{A}}=\rho_{A^{\otimes|\mathcal{L}||\mathcal{K}|}}^{\otimes |\mathcal{M}|}$\footnote{Due to the cumbersome notations we face, the tensor product states are shown for example as either $\rho_{X}^{\otimes |\mathcal{M}|}$ or $\rho_{X^{\otimes |\mathcal{M}|}}$.}.

The encoding and decoding pairs are as follows:
\begin{itemize}
\item Alice performs some encoding operation $\mathcal{E} : MLA\rightarrow A$. Let us denote the state in (\ref{Srd-state}) after channel transmission as:
\begin{align}
\nonumber
\left(\rho_{XX'(AYY')^{\otimes|\mathcal{L}||\mathcal{K}|}}\right)&^{\otimes |\mathcal{M}|-1}\\
&\otimes\rho_{XX'\left(YY'\right)^{\otimes|\mathcal{L}||\mathcal{K}|}(A)^{\otimes|\mathcal{L}||\mathcal{K}|-1}BE}^{m,(\ell,k)},
\label{resulted state}
\end{align} 
where $(m,\ell,k)\in[1:2^{r}]\times[1:2^{R}]\times[1:2^{\tilde{R}}]$ are the public message, the private message and a dummy index drawn uniformly at random by the encoder and $\rho_{XX'\left(YY'\right)^{\otimes|\mathcal{L}||\mathcal{K}|}(A)^{\otimes|\mathcal{L}||\mathcal{K}|-1}BE}^{m,(\ell,k)}$ is given by equation (\ref{srdLK}).
\begin{figure*}[!t]
\begin{align}
\label{srdLK}
\nonumber
&\rho_{XX'\left(YY'\right)^{\otimes|\mathcal{L}||\mathcal{K}|}(A)^{\otimes|\mathcal{L}||\mathcal{K}|-1}BE}^{m,(\ell,k)}\coloneqq \\
&\sum_{x}p_{X}(x)|x\rangle\langle x|_{X}\otimes|x\rangle\langle x|_{X'}\otimes \rho_{YY'A}^{x,m,(1,1)}\otimes...
\otimes \rho_{YY'A}^{x,m,(\ell,k-1)}\otimes\mathcal{N}_{A\rightarrow BE}\left(\rho_{YY'A}^{x,m,(\ell,k)}\right)
\otimes \rho_{YY'A}^{x,m,(\ell,k+1)}...\otimes \rho_{YY'A}^{x,m,(|\mathcal{L}|,|\mathcal{K}|)}.
\end{align}
\hrulefill
\end{figure*}
\item After the channel action, Bob performs a decoding operation (quantum instrument) $\mathcal{D}^{1}: BX\rightarrow \hat{M}B$ on his $\Upsilon_{T_{X}}$ systems as well as the received system, whose outputs are a classical system $\hat{M}$ and a quantum system in $\mathcal{D}(\mathcal{H}_{B})$ (the decoder will be defined formally later, see (\ref{eq_decoder1})). The action of the quantum decoder $\mathcal{D}^{1}_{BX\rightarrow \hat{M}B}$ on Bob's corresponding systems is as follows:
\begin{align}
\label{fstd}
\nonumber
\mathcal{D}^{1}_{BX\rightarrow \hat{M}B}&(\rho_{X^{\otimes|\mathcal{M}|}B}^{m,(\ell,k)})\coloneqq\\
&\sum_{m'=1}^{|\mathcal{M}|}|m'\rangle\langle m'|_{\hat{M}}\otimes\mathcal{D}^{1,m'}_{BX\rightarrow B}(\rho_{X^{\otimes|\mathcal{M}|}B}^{m,(\ell,k)}),
\end{align}
where $\{|m\rangle\}_{m=1}^{|\mathcal{M}|}$ is some orthonormal basis and $\rho_{X^{\otimes|\mathcal{M}|}B}^{m,(\ell,k)}$ can be seen from (\ref{resulted state}) by tracing out uninvolved systems. Moreover, tracing out the classical system $\hat{M}$ gives the induced quantum operation $\mathcal{D}^{1}_{BX\rightarrow B}=\sum_{m}\mathcal{D}^{1,m}_{BX\rightarrow B}$ such that its sum is trace preserving, i.e., $\text{Tr}\left\{\sum_{m'=1}^{|\mathcal{M}|}\mathcal{D}^{1,m'}_{BX\rightarrow B}(\rho_{X^{\otimes|\mathcal{M}|}B}^{m,(\ell,k)})\right\}=1$. 

Let $\sigma_{XX'(YY')^{\otimes|\mathcal{L}||\mathcal{K}|}BE}^{m,(\ell,k)}$ denote the \textit{disturbed} state after Bob applied his first decoder (this state will be defined formally later, see (\ref{srdLK-resulted})). 

\item Bob's second decoder is another quantum map $\mathcal{D}^{2}: \hat{M}BY\rightarrow \hat{L}$ which is input the classical output of the first decoder, the disturbed quantum output, Bob's $\Upsilon_{T_{Y}}$ systems and outputs a classical system $\hat{L}$.
\begin{align}
\mathcal{D}^{2}_{\hat{M}BY\rightarrow \hat{L}}(\sigma_{XY^{\otimes|\mathcal{L}||\mathcal{K}|}B}^{m,(\ell,k)})&\coloneqq\sum_{\ell=1}^{|\mathcal{L}|}p_{\hat{L}}(\ell)|\ell\rangle\langle \ell|_{\hat{L}},
\end{align}
where $\{|\ell\rangle\}_{\ell=1}^{|\mathcal{L}|}$ is some orthonormal basis and $\sigma_{XY^{\otimes|\mathcal{L}||\mathcal{K}|}B}^{m,(\ell,k)}$ comes about by tracing out uninvolved systems in the disturbed state.
\end{itemize}
Having defined the decoders, it is seen that the phrase in (\ref{e-Bob}) indicates the probability of an erroneous detection of the public message, while the expression in (\ref{p-Eve}) captures the notions of an erroneous detection of the private message as well as the secrecy condition of the eavesdropper (the latter is clarified below). After we derandomize the code, we see that the criteria mentioned in Definition \ref{maind} can be set out from these criteria by using the monotonicity of the trace distance and properly adjusting the constants.
\begin{figure*}[!t]
\begin{align}
\label{e-Bob}
P_{e}=&\{\hat{M}\neq M\}\coloneqq\frac{1}{M}\sum_{m=1}^{|\mathcal{M}|}\frac{1}{2}\left\|\mathcal{D}^{1}_{BX\rightarrow \hat{M}}(\rho_{X^{\otimes|\mathcal{M}|}B}^{m,(\ell,k)})-|m\rangle\langle m|_{\hat{M}}\right\|_{1}\leq\epsilon,\\
\label{p-Eve}
P_{priv}\coloneqq&\frac{1}{|\mathcal{L}|}\sum_{l=1}^{|\mathcal{L}|}\frac{1}{2}\left\|\mathcal{D}^{2}_{\hat{M}BY\rightarrow \hat{L}}(\sigma_{XX'(YY')^{\otimes|\mathcal{L}||\mathcal{K}|}BE}^{m,(\ell,k)})-|l\rangle\langle l|_{\hat{L}}\otimes\hat{\sigma}_{X'Y'^{\otimes|\mathcal{L}||\mathcal{K}|}E}\right\|_{1}\leq 2(\epsilon+\sqrt{\epsilon})+\sqrt{\epsilon'},\\ \nonumber
&\text{where}\\ \nonumber
&\hat{\sigma}_{X'Y'^{\otimes|\mathcal{L}||\mathcal{K}|}E}\coloneqq\sum_{x}p_{X}(x)|x\rangle\langle x|_{X'}\otimes\sigma_{Y'^{\otimes|\mathcal{L}||\mathcal{K}|}}^{x,m,(\ell,k)}\otimes\tilde{\sigma}_{E}^{x,m} \quad\text{and}\quad P(\sigma_{YE}^{x},\tilde{\sigma}_{YE}^{x})\leq\sqrt{\epsilon'}.
\end{align}
\hrulefill
\end{figure*}
\subsubsection{Correctness of Public Message: Eq. (\ref{e-Bob})}
All systems are assumed to be traced out except those used by Bob's first decoder (we could have considered multiplying those systems by identity operator as well). To decode the public message $m$, Bob employs the following decoding instrument:
\begin{align}
\label{eq_decoder1}
&\mathcal{D}^{1}_{BX\rightarrow \hat{M}B}(\rho_{X^{\otimes|\mathcal{M}|}B}^{m,(\ell,k)}) \\ \nonumber
&\hspace*{1cm} \coloneqq\sum_{m=1}^{|\mathcal{M}|}\text{Tr}\{\Lambda_{X^{|\mathcal{M}|}B}^{m}\rho_{X^{\otimes|\mathcal{M}|}B}^{m,(\ell,k)}\}|m\rangle\langle m|_{\hat{M}} \\ \nonumber
&\hspace*{3.5cm}\otimes\frac{\sqrt{\Lambda_{X^{|\mathcal{M}|}B}^{m}}\rho_{X^{\otimes|\mathcal{M}|}B}^{m,(\ell,k)}\sqrt{\Lambda_{X^{|\mathcal{M}|}B}^{m}}}{\text{Tr}\{\Lambda_{X^{|\mathcal{M}|}B}^{m}\rho_{X^{\otimes|\mathcal{M}|}B}^{m,(\ell,k)}\}},
\end{align}
where $\Lambda_{X^{|\mathcal{M}|}B}^{m}$ is given in (\ref{pubPOVM}),
\begin{figure*}[!t]
\begin{align}
\Lambda_{X^{|\mathcal{M}|}B}^{m}\coloneqq\left(\sum_{m^{\prime}=1}^{|\mathcal{M}|}\Gamma_{X^{|\mathcal{M}|}B}^{m^{\prime}}\right)^{-\frac{1}{2}}\Gamma_{X^{|\mathcal{M}|}B}^{m}\left(\sum_{m'=1}^{|\mathcal{M}|}\Gamma_{X^{|\mathcal{M}|}B}^{m'}\right)^{-\frac{1}{2}}.
\label{pubPOVM}
\end{align}
\hrulefill
\end{figure*}
and for $m\in[1:|\mathcal{M}|]$:
\begin{align*}
\Gamma^{m}_{X^{|\mathcal{M}|}B}=\mathbbm{1}_{X}^{1}\otimes\mathbbm{1}_{X}^{2}\otimes...\otimes T_{XB}^{m}\otimes...\otimes \mathbbm{1}_{X}^{|\mathcal{M}|},
\end{align*}
in which, $T_{XB}^{m}$ is a test operator distinguishing between two hypotheses, $\rho_{XB}$ and $\rho_{X}\otimes\rho_{B}$ and $\rho_{X^{\otimes|\mathcal{M}|}B}^{m,(\ell,k)}$ can be seen from (\ref{Srd-state}).
In fact, Bob needs to discriminate between the following states for different values of $m\in\mathcal{M}$
\begin{equation*}
\rho^{m,(\ell,k)}_{X^{\otimes|\mathcal{M}|}B}\coloneqq\rho_{X}^{\otimes |\mathcal{M}|-1} \otimes\rho_{XB}^{m,(\ell,k)}.
\end{equation*}                   
Note that to decode the public message $m$, Bob's decoder does not care about the copy selected by Alice among $|\mathcal{L}||\mathcal{K}|$ copies (no matter which one is selected). In other words, to accomplish the protocol for transmitting the public message, it suffices to consider $|\mathcal{M}|$ copies of $\rho_{XA}=\sum_{x}P_{X}(x)|x\rangle\langle x|\otimes\omega_{A}^{x}$ shared between Alice and Bob, where $\omega_{A}^{x}=\sum_{y}p(y|x)\omega_{A}^{x,y}$. Besides, as is clear from the former discussion, Bob's first decoder faces an $|\mathcal{M}|$-ary hypothesis testing problem. This $|\mathcal{M}|$-ary hypothesis testing problem can be reduced to a binary hypothesis testing
problem, in which a binary test operator discriminates between two hypotheses. However, it should not be confused with the fact that in general we deal with an $|\mathcal{M}|$-ary problem.

Let $T_{XB}$ be a test operator in a binary hypothesis testing scenario with null and alternative hypotheses being $\rho_{XB}$ and $\rho_{X}\otimes\rho_{B}$, respectively. Discriminator employed by Bob succeeds in guessing null and alternative hypotheses with probabilities $\textrm{Tr}\{T_{XB}\rho_{XB}\}$ and $\textrm{Tr}\{(\mathbbm{1}_{XB}-T_{XB})(\rho_{X}\otimes\rho_{B})\}$, respectively. And accordingly, the error probabilities associated to the type I and II errors are $\textrm{Tr}\{(\mathbbm{1}_{XB}-T_{XB})\rho_{XB}\}$ and $\textrm{Tr}\{T_{XB}(\rho_{X}\otimes\rho_{B})\}$, respectively.

It is notation-wise useful to assume that the error probability of the hypothesis tester is $\epsilon-\delta$ where $\delta\in (0,\epsilon)$ implying that overall probability of error ($\epsilon$) is greater than or equal to that of the hypothesis tester.
Having introduced the test operator, we can define the following measurement operator for all $m\in[1:|\mathcal{M}|]$:
\begin{align*}
\Gamma^{m}_{X^{|\mathcal{M}|}B}=\mathbbm{1}_{X}^{1}\otimes...\otimes T_{XB}^{m}\otimes...\otimes \mathbbm{1}_{X}^{|\mathcal{M}|}.
\end{align*}
 If Alice sends the $m$-th message (copy), the probability of producing the correct message at the output equals:
\begin{align}
\nonumber
\textrm{Tr}\{\Gamma^{m}_{X^{|\mathcal{M}|}B}&\rho^{m,(\ell,k)}_{X^{\otimes|\mathcal{M}|}B}\}\\ \nonumber
=\textrm{Tr}\{(\mathbbm{1}_{X}&^{1}\otimes \mathbbm{1}_{X}^{2}\otimes...\otimes T_{XB}^{m}\otimes...\otimes \mathbbm{1}_{X}^{|\mathcal{M}|})\\ \nonumber
&(\rho_{X}^{1}\otimes...\otimes\rho_{XB}^{m,(\ell,k)}\otimes...\otimes\rho_{X}^{|\mathcal{M}|})\} \\
\label{e1}
=\textrm{Tr}\{T^{m}_{XB}&\rho_{XB}^{m}\}=\textrm{Tr}\{T_{XB}\rho_{XB}\},
\end{align}
 where in the last equality we drop the dependence on $m$ since it is the same for all messages.
And probability of deciding in favor of $m'\neq m$ when $m$ was sent is equal to:
\begin{align}
\nonumber
&\textrm{Tr}\{\Gamma^{m'}_{X^{|\mathcal{M}|}B}\rho^{m,(\ell,k)}_{X^{\otimes|\mathcal{M}|}B}\}\\ \nonumber
&=\textrm{Tr}\{(\mathbbm{1}_{X}^{m}\otimes T_{XB}^{m'})(\rho_{XB}^{m,(\ell,k)}\otimes\rho_{X}^{m'})\}\\
\label{e2}
&=\textrm{Tr}\{T_{XB}^{m'}(\rho_{B}^{m,(\ell,k)}\otimes\rho_{X}^{m'})\}=\textrm{Tr}\{T_{XB}(\rho_{B}\otimes\rho_{X})\},
\end{align}
where in the last equality we remove the index $m'$ because this quantity is the same for all $m'\ne m$. This endorses our claim saying that we are facing a binary hypothesis testing problem. From the aforementioned measurement operators, the square-root measurement given in (\ref{pubPOVM}) is formed acting as Bob's POVM to detect the public message $m$.
The mentioned POVM construction and the coding scheme, known as position-based coding, first appeared in \cite{ADJ17} and \cite{AJW17}.

We now focus on the analysis of the error probability of the position-based decoder. The POVM elements above are unitary permutations of one another. In particular, it can be easily shown that all of the elements can be reached by a unitary permutation of the first one, i.e., $\Lambda_{X^{|\mathcal{M}|}B}^{m}=U_{X^{|\mathcal{M}|}B}^{\pi(m)}\Lambda_{X^{|\mathcal{M}|}B}^{1}U_{X^{|\mathcal{M}|}B}^{\pi(m)\dagger}$ in which $\pi(.)$ denotes the permutatin operator \cite{Wil17}. Having said this, we find the probability of error for the first message, i.e., Alice received $m=1$ and has chosen and sent one of the $|\mathcal{L}||\mathcal{K}|$ $A$ subsystems of the first copy over the channel. We emphasize again that although Alice selects a particular $A$ subsystem out of $|\mathcal{L}||\mathcal{K}|$ copies based on reliability and security of the private message, at this point, when Bob aims to estimate the public message, no matter which $A$ was chosen by Alice, it does not affect Bob's decision about the public message.

 We begin by applying the Hayashi-Nagaoka operator inequality (Lemma \ref{HN}) with $S=\Gamma_{X^{|\mathcal{M}|}B}^{1}$ and $T=\sum_{m\neq 1}\Gamma_{X^{|\mathcal{M}|}B}^{m}$ (This $T$ should not be confused with the test operator $T_{XB}^{m}$):
\begin{align*}
Pr&(\hat{M}\neq1|M=1)\\
&=\textrm{Tr}\{(\mathbbm{1}_{X^{|\mathcal{M}|}B}-\Lambda_{X^{|\mathcal{M}|}B}^{1})\rho_{X^{\otimes|\mathcal{M}|}B}^{1,(\ell,k)}\} \\
&\leq\textrm{Tr}\{((1+c)(\mathbbm{1}_{X^{|\mathcal{M}|}B}-\Gamma_{X^{|\mathcal{M}|}B}^{1})\rho_{X^{\otimes|\mathcal{M}|}B}^{1,(\ell,k)}\}\\
&\hspace*{1.5cm}+(2+c+c^{-1})\textrm{Tr}\{(\sum_{m\neq 1}\Gamma_{X^{|\mathcal{M}|}B}^{m})\rho_{X^{\otimes|\mathcal{M}|}B}^{1,(\ell,k)}\}\\
&=(1+c)\textrm{Tr}\{(\mathbbm{1}_{X^{|\mathcal{M}|}B}-\Gamma_{X^{|\mathcal{M}|}B}^{1})\rho_{X^{\otimes|\mathcal{M}|}B}^{1,(\ell,k)}\}\\
&\hspace*{1.5cm}+(2+c+c^{-1})\textrm{Tr}\{(\sum_{m\neq 1}\Gamma_{X^{|\mathcal{M}|}B}^{m})\rho_{X^{\otimes|\mathcal{M}|}B}^{1,(\ell,k)}\}
\end{align*}
\begin{align*}
&=(1+c)\textrm{Tr}\{(\mathbbm{1}_{X}^{1}-T_{XB}^{1})\rho_{XB}^{1,(\ell,k)}\}\\
&\hspace*{1.5cm}+(2+c+c^{-1})\sum_{m\neq1}\textrm{Tr}\{T_{XB}^{m}(\rho_{B}^{m,(\ell,k)}\otimes \rho_{X}^{m})\} \\
&=(1+c)\textrm{Tr}\{(\mathbbm{1}_{XB}-T_{XB})\rho_{XB}\}\\
&\hspace*{1.5cm}+(2+c+c^{-1})(|\mathcal{M}|-1)\textrm{Tr}\{T_{XB}(\rho_{B}\otimes \rho_{X})\},
\end{align*}
where in the second last equality, the first and second terms follow from (\ref{e1}) and (\ref{e2}), respectively. Let $\Pi_{XB}$ be the optimal test operator in the following optimization: (see Definition \ref{htreo} and  Definition \ref{htma})
\begin{align*}
I_{H}^{\epsilon-\delta}(X;B)_{\rho_{XB}}\coloneqq-\log_{2}\inf_{\substack{0\leq T_{XB}\leq \mathbbm{1}, \\ \alpha(T_{XB}, \rho_{XB})\leq\epsilon-\delta}}\beta(T_{XB}, \rho_{X}\otimes\rho_{B}),
\end{align*}
then,
\begin{align*}
\textrm{Tr}\{&(\mathbbm{1}_{X^{|\mathcal{M}|}B}-\Lambda_{X^{|\mathcal{M}|}B}^{1})\rho_{X^{\otimes|\mathcal{M}|}B}^{1,(\ell,k)}\} \\
&\leq(1+c)\textrm{Tr}\{(\mathbbm{1}_{XB}-\Pi_{XB})\rho_{XB}\}\\
&\hspace*{1.5cm}+(2+c+c^{-1})(|\mathcal{M}|-1)\textrm{Tr}\{\Pi_{XB}(\rho_{B}\otimes \rho_{X})\} \\
&\leq(1+c)(\epsilon-\delta) +(2+c+c^{-1})|\mathcal{M}|2^{-I_{H}^{\epsilon-\delta}(X;B)_{\rho}}.
\end{align*}
The last term above is set equal to $\epsilon$, if we solve for $|\mathcal{M}|$, we end up with the following term
\begin{align*}
\log_{2}|\mathcal{M}|=I_{H}^{\epsilon-\delta}(X;B)_{\rho}+\log_{2}\left(\frac{\epsilon-(1+c)(\epsilon-\delta)}{2+c+c^{-1}}\right),
\end{align*} 
the expression inside the $\log$ has a global maximum with respect to $c$, i.e., the parabola is down-side. We put first derivative equal to zero and pick $c=\frac{\delta}{2\epsilon-\delta}$ and by doing so finally the following bound holds:
\begin{equation}
\log_{2}|\mathcal{M}|\leq I_{H}^{\epsilon-\delta}(X;B)_{\rho}-\log_{2}(\frac{4\epsilon}{\delta^{2}}),
\end{equation}
and average probability of error of the public message for the one-shot assisted code is
\begin{align}
\frac{1}{|\mathcal{M}|}\sum_{m=1}^{|\mathcal{M}|}\textrm{Tr}\{(\mathbbm{1}_{X^{|\mathcal{M}|}B}-\Lambda_{X^{|\mathcal{M}|}B}^{m})\rho_{X^{\otimes|\mathcal{M}|}B}^{m,(\ell,k)}\}\leq\epsilon.
\end{align}

In the following, we deal with the private message and the second decoder. Before we move on to the privacy analysis, we make a couple of remarks. If the first decoder fails, the second decoder breaks down completely since as is intuitively clear, it ends up with a state having zero information about the position of the sent message. We precisely evaluate the contribution of the first decoder to the error of the second decoder. Moreover, Bob's first decoder acts on his $X$ systems as well as the output of the channel. The $Y$ systems remain intact and in fact, when Bob applies the first decoder, one can assume that the uninvolved systems are being multiplied by the identity operators. Considering this point and the action of the POVM, the resulting state on systems $X$ and $B$ are (up to normalization) $\sqrt{\Lambda_{X^{|\mathcal{M}|}B}^{m}}\rho_{X^{\otimes|\mathcal{M}|}B}^{m,(\ell,k)}\sqrt{\Lambda_{X^{|\mathcal{M}|}B}^{m}}$, and by taking uninvolved systems into account, we define the state that passes to the second decoder as in (\ref{srdLK-resulted}).
\begin{figure*}[!t]
\begin{align}
\label{srdLK-resulted}
\nonumber
&\sigma_{XX'\left(YY'\right)^{\otimes|\mathcal{L}||\mathcal{K}|}BE}^{m,(\ell,k)}\\
&\hspace{1cm}\coloneqq\sum_{x}p_{X}(x)|x\rangle\langle x|_{X}\otimes|x\rangle\langle x|_{X'}\otimes\sigma_{YY'}^{x,m,(1,1)}\otimes...\otimes
 \sigma_{YY'}^{x,m,(\ell,k-1)}\otimes\sigma_{YY'BE}^{x,m,(\ell,k)}\otimes \sigma_{YY'}^{x,m,(\ell,k+1)}...\otimes \sigma_{YY'}^{x,m,(|\mathcal{L}|,|\mathcal{K}|)}.
\end{align}
\hrulefill
\end{figure*}

\subsubsection{Correctness and secrecy of Private Message, (Privacy error) Eq. (\ref{p-Eve})} 
Reconsider the state in (\ref{srdLK-resulted}) showing the state resulted from transmitting the $(\ell,k)$-th $A$ subsystem through the channel (for a given $m$) after Bob applies his first decoder. Remember that in the first part of the protocol it did not matter which copy out of $|\mathcal{L}||\mathcal{K}|$ copies was chosen but now it does matter as Bob and Eve try to decode the private message.
Bob's decoder for the private message $\ell$ is constructed as follows:
\begin{align}
\label{eq_decoder2}
\mathcal{D}^{2}_{\hat{M}BY\rightarrow \hat{L}}&(\sigma_{XY^{\otimes|\mathcal{L}||\mathcal{K}|}B}^{m,(\ell,k)}) \\ \nonumber
&\coloneqq\sum_{l=1}^{|\mathcal{L}|}\text{Tr}\{P_{XY^{|\mathcal{L}||\mathcal{K}|}B}^{\ell}\sigma_{XY^{\otimes|\mathcal{L}||\mathcal{K}|}B}^{m,(\ell,k)}\}|\ell\rangle\langle \ell|_{\hat{L}},
\end{align}
where for all $x\in\mathcal{X}$
\begin{align*}
&P_{XY^{|\mathcal{L}||\mathcal{K}|}B}^{\ell}=\sum_{k=1}^{|\mathcal{K}|}P_{XY^{|\mathcal{L}||\mathcal{K}|}B}^{(\ell,k)},
\end{align*}
and $P_{XY^{|\mathcal{L}||\mathcal{K}|}B}^{x,(\ell,k)}$ is given in (\ref{priPOVM}),
\begin{figure*}[!t]
\begin{align}
\label{priPOVM}
P_{XY^{|\mathcal{L}||\mathcal{K}|}B}^{(\ell,k)}\coloneqq
\left(\sum_{\ell'=1}^{|\mathcal{L}|}\sum_{k'=1}^{|\mathcal{K}|}N_{XY^{|\mathcal{L}||\mathcal{K}|}B}^{(\ell',k')}\right)^{-\frac{1}{2}}N_{XY^{|\mathcal{L}||\mathcal{K}|}B}^{(\ell,k)}\left(\sum_{\ell'=1}^{|\mathcal{L}|}\sum_{k'=1}^{|\mathcal{K}|}N_{XY^{|\mathcal{L}||\mathcal{K}|}B}^{(\ell',k')}\right)^{-\frac{1}{2}},
\end{align}
where for all $\ell \in [1:|\mathcal{L}|], \textrm{and } k \in [1:|\mathcal{K}|],$
\begin{align}
\label{dn1}
 N_{XY^{|\mathcal{L}||\mathcal{K}|}B}^{(\ell,k)}\coloneqq|x\rangle\langle x|_{X}\otimes\mathbbm{1}_{Y}^{(1,1)} \otimes...\otimes  \mathbbm{1}_{Y}^{(1,|\mathcal{K}|)} \otimes...\otimes  \mathbbm{1}_{Y}^{(\ell,k-1)}\otimes Z_{YB}^{(\ell,k)}\otimes \mathbbm{1}_{Y}^{(\ell,k+1)}...\otimes  \mathbbm{1}_{Y}^{(|\mathcal{L}|,|\mathcal{K}|)}.
\end{align}
\hrulefill
\end{figure*}
in which, for all $x\in\mathcal{X}$, $Z_{YB}$ is a binary test operator distinguishing between  two hypotheses $\sigma_{YB}^{x}$ and $\sigma_{Y}^{x}\otimes\sigma_{B}^{x}$ with an error of $\epsilon-\delta$, i.e.,  
 \begin{align*}
  \text{Tr}\{Z_{YB}\sigma_{YB}^{x}\}\geq 1-(\epsilon-\delta),
 \end{align*}
 where $\epsilon\in(0, 1)$ and $\delta\in(0, \epsilon)$. Note that the variable $x$ appearing in the operator indicates the fact that the decoding works for all $x\in\mathcal{X}$.
 
Bob has to be able to distinguish between states $\sigma_{XY^{\otimes|\mathcal{L}||\mathcal{K}|}B}^{m,(1,1)}, \sigma_{XY^{\otimes|\mathcal{L}||\mathcal{K}|}B}^{m,(1,2)},..., \sigma_{XY^{\otimes|\mathcal{L}||\mathcal{K}|}B}^{m,(l,k)}, \sigma_{XY^{\otimes|\mathcal{L}||\mathcal{K}|}B}^{m,(|\mathcal{L}|,|\mathcal{K}|)}$; We will see that this amounts to Bob being able to distinguish between the following states: 
\begin{align*}
&\sum_{x}p_{X}(x)|x\rangle\langle x|_{X}\otimes\sigma_{YB}^{x},\\
&\sum_{x}p_{X}(x)|x\rangle\langle x|_{X}\otimes\sigma_{Y}^{x}\otimes\sigma_{B}^{x}, 
\end{align*}
or more precisely, between state $\sigma_{YB}^{x}$ and $\sigma_{Y}^{x}\otimes \sigma_{B}^{x}$ for all $x\in\mathcal{X}$. We importantly note that after detecting the public message $m$, Bob is faced a $|\mathcal{L}||\mathcal{K}|$-ary hypothesis testing problem. This scenario should not be confused by the binary hypothesis testing above, i.e., Alice distinguishes between $\sigma_{YB}^{x}$ and $\sigma_{Y}^{x}\otimes \sigma_{B}^{x}$ for all $x\in\mathcal{X}$, the latter happens to be a byproduct of the general scenario once we go into the error analysis.   
Now see that if the pair $(\ell,k)$ was chosen, the action of the operator $N_{Y^{|\mathcal{L}||\mathcal{K}|}B}^{(\ell,k)}$ would be as follows:
\begin{align*}
\textrm{Tr}&\{N_{XY^{|\mathcal{L}||\mathcal{K}|}B}^{(\ell,k)}\sigma_{XY^{\otimes|\mathcal{L}||\mathcal{K}|}B}^{m,(\ell,k)}\} \\
&\hspace*{2cm}=\sum_{x}p_{X}(x)\textrm{Tr}\{Z_{YB}^{(\ell,k)}\sigma_{YB}^{x,m,(\ell,k)}\},
\end{align*}
and for any other operator, i.e., the private message-local key pair $(\ell,k)$ is confused by $(\ell',k')$, either $k\neq k'$, $l\neq \ell'$ or $(k\neq k', \ell\neq \ell')$:
\begin{align*}
&\textrm{Tr}\{N_{XY^{|\mathcal{L}||\mathcal{K}|}B}^{(\ell',k')}\sigma_{XY^{\otimes|\mathcal{L}||\mathcal{K}|}B}^{m,(\ell,k)}\}\\
&=\textrm{Tr}\bigg\{|x\rangle\langle x|_{X}\otimes(Z_{YB}^{{(\ell',k')}}\otimes \mathbbm{1}_{Y}^{(\ell,k)})\\
&\hspace*{1cm}(\sum_{x}p_{X}(x)|x\rangle\langle x|_{X}\otimes\sigma_{Y}^{x,m,(\ell',k')}\otimes \sigma_{YB}^{x,m,(\ell,k)})\bigg\}\\
&=\sum_{x}p_{X}(x)\textrm{Tr}\{Z_{YB}^{(\ell',k')}(\sigma_{Y}^{x,m,(\ell',k')}\otimes \sigma_{B}^{x,m,(\ell,k)})\}.
\end{align*}
We can think of the states $\sigma_{YB}^{x,m}$ and $\sigma_{Y}^{x,m}\otimes \sigma_{B}^{x,m}$ as the null and alternative hypotheses, respectively. As a typical procedure in quantum error analysis, Bob forms the square-root measurement operators given in (\ref{priPOVM}) acting as his POVMs to detect the private message-local key pair $(\ell,k)$. 
It can be shown that each measurement operator $P_{Y^{|\mathcal{L}||\mathcal{K}|}B}^{(\ell,k)}$ is related to the first one $P_{Y^{|\mathcal{L}||\mathcal{K}|}B}^{(1,1)}$ by a unitary permutation of $Y^{|\mathcal{L}||\mathcal{K}|}$ systems for all $x\in\mathcal{X}$. This fact gives rise to the following identity,
for all $\ell \in [1:|\mathcal{L}|]$ and $k\in [1:|\mathcal{K}|]$:
\begin{align*}
 \textrm{Tr}\{&(\mathbbm{1}_{XY^{|\mathcal{L}||\mathcal{K}|}B}-P_{XY^{|\mathcal{L}||\mathcal{K}|}B}^{(1,1)})\sigma_{XY^{\otimes|\mathcal{L}||\mathcal{K}|}B}^{m,(1,1)}\}\\
 &\hspace*{1.5cm}=\textrm{Tr}\{(\mathbbm{1}_{XY^{|\mathcal{L}||\mathcal{K}|}B}-P_{XY^{|\mathcal{L}||\mathcal{K}|}B}^{(\ell,k)})\sigma_{XY^{\otimes|\mathcal{L}||\mathcal{K}|}B}^{m,(\ell,k)}\},
\end{align*}
meaning that the error probability is the same for all private messages, in other words, it is independent from a particular chosen twin $(\ell,k)$; And again this implies that average error probability equals individual error probabilities. In what follows, we deploy Hayashi-Nagaoka operator inequality (Lemma \ref{HN}) to analyze the error probability. Let's assume $(\ell=1,k=1)$ was sent. Moreover, let's choose $S=N_{Y^{|\mathcal{L}||\mathcal{K}|}B}^{(1,1)}$ and $T=\sum_{\ell'\ne 1}\sum_{k'\ne1}N_{Y^{|\mathcal{L}||\mathcal{K}|}B}^{(\ell',k')}$ in Hayashi-Nagaoka inequality. We have
\begin{align*}
&\textrm{Tr}\{(\mathbbm{1}_{Y^{|\mathcal{L}||\mathcal{K}|}B}-P_{XY^{|\mathcal{L}||\mathcal{K}|}B}^{(1,1)})\sigma_{XY^{\otimes|\mathcal{L}||\mathcal{K}|}B}^{m,(1,1)}\} \\
&\leq (1+c)\textrm{Tr}\{(\mathbbm{1}_{Y^{|\mathcal{L}||\mathcal{K}|}B}-N_{XY^{|\mathcal{L}||\mathcal{K}|}B}^{(1,1)})\sigma_{XY^{\otimes|\mathcal{L}||\mathcal{K}|}B}^{m,(1,1)}\}\\
&\hspace*{1cm}+(2+c+c^{-1})\sum_{l'\ne1}\sum_{k'\ne1} \textrm{Tr}\{N_{XY^{|\mathcal{L}||\mathcal{K}|}B}^{(\ell',k')}\sigma_{XY^{\otimes|\mathcal{L}||\mathcal{K}|}}^{m,(1,1)}\} \\
&=(1+c)\sum_{x}p_{X}(x)\textrm{Tr}\{(\mathbbm{1}_{YB}-Z_{YB}^{(1,1)})\sigma_{YB}^{x,m,(1,1)}\}\\
&+(2+c+c^{-1})\sum_{x}p_{X}(x)\big(\\
&\hspace*{1.5cm}\sum_{l'\ne1}\sum_{k'\ne1}\textrm{Tr}\{Z_{YB}^{(\ell',k')}(\sigma_{Y}^{x,m,(\ell',k')}\otimes\sigma_{B}^{x,m,(1,1)})\}\big) \\
&=(c+1)\sum_{x}p_{X}(x)\textrm{Tr}\{(\mathbbm{1}_{YB}-Z_{YB})\sigma_{YB}^{x,m}\}\\
&+(2+c+c^{-1})(|\mathcal{L}||\mathcal{K}|-1)(\\
&\hspace*{3cm}\sum_{x}p_{X}(x)\textrm{Tr}\{Z_{YB}(\sigma_{Y}^{x,m}\otimes \sigma_{B}^{x,m})\})
\end{align*}

For each realization $x$, let $\Theta_{YB}^{x}$ denote the measurement operator that is the answer to the optimization mentioned in Definition \ref{htreo} with $\alpha(Z_{YB}, \sigma_{YB}^{x})\coloneqq\text{Tr}\{(\mathbbm{1}-Z_{YB})\sigma_{YB}^{x}\}$ and $\beta(Z_{YB}, \sigma_{Y}^{x}\otimes\sigma_{B}^{x})\coloneqq\text{Tr}\{Z_{YB}(\sigma_{Y}^{x}\otimes\sigma_{B}^{x})\}$ where by assumption it detects the joint state with an error probability of $\epsilon-\delta$ where $\delta\in (0,\epsilon)$. This optimization can be done for all $x$, but from the definition of the conditional hypothesis testing mutual information (Definition \ref{chtma}), the $x$ minimizing the expression given in equation (\ref{opt1}) over a nearby distribution is of particular interest in error analysis;
\begin{figure*}[!t]
\begin{align}
\label{opt1}
I_{H}^{\epsilon-\delta}(Y;B|X)_{\sigma_{XYB}}\coloneqq
\max_{\substack{\sigma'_{X}}} \min_{\substack{x\in\text{supp}\left(\sigma'_{X}\right)}}\left\{-\log_{2}\inf_{\substack{0\leq Z_{YB}^{x}\leq \mathbbm{1}, \\ \alpha(Z_{YB}^{x}, \sigma_{YB}^{x})\leq\epsilon-\delta}}\beta(Z_{YB}^{x}, \sigma_{Y}^{x}\otimes\sigma_{B}^{x})\right\},
\end{align}
\begin{align*}
\text{where}\qquad \sigma_{XYB}=\sum_{x}p_{X}(x)|x\rangle\langle x|_{X}\otimes\sigma_{YB}^{x}\qquad\text{and}\qquad P(\sigma'_{X}, \sigma_{X})\leq\epsilon''.
\end{align*}
\hrulefill
\end{figure*}
The error probability simplifies as follows: 
\begin{align*}
\textrm{Tr}&\{(\mathbbm{1}_{Y^{|\mathcal{L}||\mathcal{K}|}B}-P_{XY^{|\mathcal{L}||\mathcal{K}|}B}^{(1,1)})\sigma_{XY^{\otimes|\mathcal{L}||\mathcal{K}|}B}^{m,(1,1)}\} \nonumber \\
&\leq (c+1)\sum_{x}p_{X}(x)\textrm{Tr}\{(\mathbbm{1}_{YB}-\Theta_{YB})\sigma_{YB}^{x,m}\}\\
&+(2+c+c^{-1})|\mathcal{L}||\mathcal{K}|\sum_{x}p_{X}(x)\textrm{Tr}\{\Theta_{YB}(\sigma_{Y}^{x,m}\otimes \sigma_{B}^{x,m})\} \nonumber \\
&\leq (c+1)(\epsilon-\delta)\\
&\hspace*{1.5cm}+(2+c+c^{-1})|\mathcal{L}||\mathcal{K}|2^{-I_{H}^{\epsilon-\delta}\left(Y;B|X\right)_{\sigma_{XYB}}},
\end{align*}
where in the last line, the first expression is derived from the assumption that for all $x$, $\text{Tr}\{\Theta_{YB}\sigma_{YB}^{x,m}\}\geq 1-(\epsilon-\delta)$, and the second expression follows from (\ref{opt1}).
By putting the last line above equal to $\epsilon$ (Bob's error in detecting private message is $\epsilon$) and solving it for $|\mathcal{L}||\mathcal{K}|$, we get: 
\begin{align*}
\log_{2}|\mathcal{L}||\mathcal{K}|=&I_{H}^{\epsilon-\delta}\left(Y;B|X\right)_{\sigma_{XYB}}\\
&\hspace*{2cm}+\log_{2}\left(\frac{\epsilon-(1+c)(\epsilon-\delta)}{2+c+c^{-1}}\right).
\end{align*}
The right-hand side of the expression above should be maximized with respect to $c$. Since it is a down-side parabola when it comes to maximization, we pick its global maximum which occurs at $c=\frac{\delta}{2\epsilon-\delta}$. By plugging it back into the expression we end up having:
\begin{align*}
\log_{2}|\mathcal{L}||\mathcal{K}|\leq I_{H}^{\epsilon-\delta}\left(Y;B|X\right)_{\sigma_{XYB}}-\log_{2}(\frac{4\epsilon}{\delta^{2}}).
\end{align*}
The derivation above ensures that in the privacy error in (\ref{p-Eve}), Bob's error in detecting private message is satisfied (note that each separate criterion comes about by tracing out the other one).

We now turn our attention to Eve's state and security criterion which is merged into (\ref{p-Eve}). We also assume that Eve has detected the public message. From (\ref{srdLK-resulted}), for a fixed $(\ell,k)$, Eve's state is\footnote{Note that the state in (\ref{srdLK-resulted}) denotes the disturbed state after Bob finds the public message, without loss of generality, we also assume Eve affects the initial state in the same way.}
\begin{align*}
\sigma_{X'Y^{'\otimes|\mathcal{L}||\mathcal{K}|}E}^{m,(\ell,k)}&=\sum_{x}p_{X}(x)|x\rangle\langle x|_{X'}\otimes\sigma_{Y'}^{x,m,(1,1)}\otimes...\\
&\hspace*{1cm}\otimes\sigma_{Y'E}^{x,m,(\ell,k)}\otimes...\otimes\sigma_{Y'}^{x,m,(|\mathcal{L}|,|\mathcal{K}|)}.
\end{align*}
As we discussed before, $k$ is a local key exclusively in possession of Alice and for a given private message $\ell$, it is chosen uniformly at random; Hence, for a given message $\ell$, the state of Eve can be written as equation (\ref{E1}).
\begin{figure*}[!t]
\begin{align}
\label{E1}
\nonumber
\sigma_{X'Y^{'\otimes|\mathcal{L}||\mathcal{K}|}E}^{m,\ell}\coloneqq&\frac{1}{|\mathcal{K}|}\sum_{k=1}^{|\mathcal{K}|}\sigma_{X'Y^{'\otimes|\mathcal{L}||\mathcal{K}|}E}^{m,(\ell,k)}\coloneqq\sum_{x}p_{X}(x)|x\rangle\langle x|_{X'}\otimes\sigma_{Y'}^{x,m,(1,1)}\otimes...\\
&\otimes\left[\frac{1}{|\mathcal{K}|}\sum_{k=1}^{|\mathcal{K}|}\sigma_{Y'}^{x,m,(\ell,1)}\otimes...\otimes\sigma_{Y'E}^{x,m,(\ell,k)}\otimes...\otimes\sigma_{Y'}^{x,m,(\ell,|\mathcal{K}|)}\right]\otimes...\otimes\sigma_{Y'}^{x,m,(|\mathcal{L}|,|\mathcal{K}|)}.
\end{align}
\hrulefill
\end{figure*}
We would like her to learn almost nothing about the sent private message. In other words, her state becomes independent from the chosen index $\ell$:

\begin{align}
\label{s-Eve}
\forall m ,\ell :\quad
 \frac{1}{2}\|\sigma_{X'Y^{'\otimes|\mathcal{L}||\mathcal{K}|}E}^{m,\ell}-\hat{\sigma}_{X'Y'^{\otimes|\mathcal{L}||\mathcal{K}|}E}\|_{1}\leq \sqrt{\epsilon'},
\end{align}
where 
\begin{align}
\label{Evelearn}
\hat{\sigma}_{X'Y^{'\otimes|\mathcal{L}||\mathcal{K}|}E}\coloneqq\sum_{x}p_{X}(x)|x\rangle\langle x|_{X'}\otimes\sigma_{Y'^{\otimes|\mathcal{L}||\mathcal{K}|}}^{x,m}\otimes\tilde{\sigma}_{E}^{x,m}
\end{align}
 for $\epsilon'\in(0,1)$ and some state $\tilde{\sigma}_{E}^{m,x}$ that is the marginal of $\tilde{\sigma}_{Y'E}^{m,x}$ and $P(\sigma_{Y'E}^{m,x},\tilde{\sigma}_{Y'E}^{m,x})\leq\sqrt{\epsilon'}-\delta'$ in which $\delta'\in(0,\sqrt{\epsilon'})$. From the invariance of trace distance with respect to tensor-product states, we can expand the security constraint (\ref{s-Eve}) as given by (\ref{figur1}).
\begin{figure*}[!t]
 \begin{align}
 \label{figur1}
 \nonumber
 &\frac{1}{2}\left\|\sigma_{X'Y^{'\otimes|\mathcal{L}||\mathcal{K}|}E}^{m,\ell}-\sum_{x}p_{X}(x)|x\rangle\langle x|_{X'}\otimes\sigma_{Y^{'\otimes|\mathcal{L}||\mathcal{K}|}}^{x,m}\otimes\tilde{\sigma}_{E}^{x,m}\right\|_{1}\\ \nonumber
&=\frac{1}{2}\left\|\sum_{x}p_{X}(x)|x\rangle\langle x|_{X'}\otimes\left(\sigma_{Y^{'\otimes|\mathcal{L}||\mathcal{K}|}E}^{x,m,\ell}-\sigma_{Y^{'\otimes|\mathcal{L}||\mathcal{K}|}}^{x,m}\otimes\tilde{\sigma}_{E}^{x,m}\right)\right\|_{1}=\sum_{x}p_{X}(x)\frac{1}{2}\left\|\sigma_{Y^{'\otimes|\mathcal{L}||\mathcal{K}|}E}^{x,m,\ell}-\sigma_{Y^{'\otimes|\mathcal{L}||\mathcal{K}|}}^{x,m}\otimes\tilde{\sigma}_{E}^{x,m}\right\|_{1}\\
&=\frac{1}{2}\sum_{x}p_{X}(x)\left\|\frac{1}{K}\sum_{k=1}^{|\mathcal{K}|}\sigma_{Y'}^{x,m,(\ell,1)}\otimes...\otimes\sigma_{Y'}^{x,m,(\ell,k-1)}\otimes\left(\sigma_{Y'E}^{x,m,(\ell,k)}-\sigma_{Y'}^{x,m}\otimes\tilde{\sigma}_{E}^{x,m}\right)\otimes
\sigma_{Y'}^{x,m,(\ell,k+1)}\otimes...\otimes \sigma_{Y'}^{x,m,(|\mathcal{L}|,|\mathcal{K}|)}\right\|_{1}.
\end{align}
\hrulefill
\end{figure*}

From the convex-split lemma and the definition of the conditional smooth max-mutual information (see Definition \ref{casmma}), if the following condition holds\footnote{To maintain consistency, in the following expression, we show Eve's $X'$ and $Y'$ systems with $X$ and $Y$, respectively.},
\begin{align}
\log_{2}{|\mathcal{K}|}=\tilde{I}_{max}^{\sqrt{\epsilon'}-\delta'}(E;Y|X)_{\sigma}+2\log_{2}(\frac{1}{\delta'}),
\end{align}
then
\begin{align*}
 P(\sigma_{X'Y^{'\otimes|\mathcal{L}||\mathcal{K}|}E}^{m,\ell}, \hat{\sigma}_{X'Y^{'\otimes|\mathcal{L}||\mathcal{K}|}E})\leq\sqrt{\epsilon'}
\end{align*}
is satisfied with $\hat{\sigma}_{X'Y'^{|\mathcal{L}||\mathcal{K}|}E}$ defined in (\ref{Evelearn}) and from the relation between purified distance and trace distance correctness of (\ref{s-Eve}) is guaranteed. Note also that $P(\sigma_{E}^{x,m}, \tilde{\sigma}_{E}^{x,m})\leq P(\sigma_{YE}^{x,m}, \tilde{\sigma}_{YE}^{x,m})\leq \sqrt{\epsilon'}-\delta'$.
So far, we have shown the correctness of two separate criteria for the assisted code. For our purposes here we would like to have a single condition for the private message encompassing both conditions discussed lately and so in the following, by sticking to the reciepe set out by \cite{Wil17}, we try to merge two conditions and deal with a single \textit{privacy error}. We see that the single criterion will be beneficial once we derandomize the code and upon derandomization, the requirements set out in the definition of the unassisted code will be fulfilled.
 
We saw that the average error probability is equal to the individual error probabilities:
\begin{align}
\label{eqai}
\nonumber
&\textrm{Tr}\{(\mathbbm{1}_{XY^{|\mathcal{L}||\mathcal{K}|}B}-P_{XY^{|\mathcal{L}||\mathcal{K}|}B}^{(\ell,k)})\sigma_{XY^{\otimes|\mathcal{L}||\mathcal{K}|}B}^{m,(\ell,k)}\}=\\
&\frac{1}{|\mathcal{L}||\mathcal{K}|}\sum_{l=1}^{|\mathcal{L}|}\sum_{k=1}^{|\mathcal{K}|}\textrm{Tr}\{(\mathbbm{1}_{Y^{|\mathcal{L}||\mathcal{K}|}B}-P_{XY^{|\mathcal{L}||\mathcal{K}|}B}^{(\ell,k)})\sigma_{XY^{\otimes|\mathcal{L}||\mathcal{K}|}B}^{m,(\ell,k)}\}\leq \epsilon
\end{align}
We continue by expanding $\sigma_{XY^{\otimes|\mathcal{L}||\mathcal{K}|}B}^{m,(\ell,k)}=\sum_{x}p_{X}(x)|x\rangle\langle x|_{X}\otimes\sigma_{Y^{\otimes|\mathcal{L}||\mathcal{K}|}B}^{x,m,(\ell,k)}$ as in equation (\ref{scdstate}).
\begin{figure*}[!t]
\begin{align*}
\sigma_{Y^{\otimes|\mathcal{L}||\mathcal{K}|}B}^{x,m,(\ell,k)}
&=\sigma_{Y}^{x,m,(1,1)}\otimes...\otimes\sigma_{Y}^{x,m,(1,|\mathcal{K}|)}\otimes...\otimes\sigma_{Y}^{x,m,(\ell,k-1)}\otimes\sigma_{YB}^{x,m,(\ell,k)}\otimes\sigma_{Y}^{x,m,(\ell,k+1)}...\otimes\sigma_{Y}^{x,m,(|\mathcal{L}|,|\mathcal{K}|)} \nonumber \\
&=\sum_{y_{11}}p(y_{11}|x)|y_{11}\rangle\langle y_{11}|\otimes...\otimes\sum_{y_{1|\mathcal{K}|}}p(y_{1|\mathcal{K}|}|x)|y_{1|\mathcal{K}|}\rangle\langle y_{1|\mathcal{K}|}|\otimes...\otimes\sum_{y_{\ell k-1}}p(y_{\ell k-1})|y_{\ell k-1}\rangle\langle y_{\ell k-1}| \nonumber \\    
&\hspace{.4cm}\otimes...\otimes\sum_{y_{\ell k}}p(y_{\ell k}|x)|y_{\ell k}\rangle\langle y_{\ell k}|\otimes\sigma_{B}^{x,m,y_{\ell k}} \otimes\sum_{y_{\ell k+1}}p(y_{\ell k+1}|x)|y_{\ell k+1}\rangle\langle y_{\ell k+1}|...\otimes\sum_{y_{|\mathcal{L}||\mathcal{K}|}}p(y_{|\mathcal{L}||\mathcal{K}|}|x)|y_{|\mathcal{L}||\mathcal{K}|}\rangle\langle y_{|\mathcal{L}||\mathcal{K}|}| \nonumber \\ \nonumber
&=\sum_{y_{11}y_{12}...y_{lk}...y_{|\mathcal{L}||\mathcal{K}|}}p(y_{11}|x)p(y_{12}|x)...p(y_{\ell k}|x)...p(y_{|\mathcal{L}||\mathcal{K}|}|x)|y_{11}...y_{\ell k}...y_{|\mathcal{L}||\mathcal{K}|}\rangle\langle y_{11}...y_{\ell k}...y_{|\mathcal{L}||\mathcal{K}|}|\otimes\sigma_{B}^{x,m,y_{\ell k}}, 
\end{align*}
hence
\begin{align}
\sigma_{XY^{\otimes|\mathcal{L}||\mathcal{K}|}B}^{m,(\ell,k)}=\sum_{x,y_{11}y_{12}...y_{lk}...y_{|\mathcal{L}||\mathcal{K}|}}p_{XY^{|\mathcal{L}||\mathcal{K}|}}(x,y_{11}...y_{lk}...y_{|\mathcal{L}||\mathcal{K}|})|x\rangle\langle x|_{X}\otimes |y_{11}...y_{\ell k}...y_{|\mathcal{L}||\mathcal{K}|}\rangle\langle y_{11}...y_{\ell k}...y_{|\mathcal{L}||\mathcal{K}|}|\otimes\sigma_{B}^{x,m,y_{\ell k}}.
\label{scdstate}
\end{align}
\hrulefill
\end{figure*}

Reconsider the optimal test operator $\Theta_{YB}$, we can write the following equation:

\begin{align*}
\textrm{Tr}\{\Theta_{YB}\sigma_{YB}^{x}\}
&=\textrm{Tr}\{\Theta_{YB}(\sum_{y}p(y|x)|y\rangle\langle y|\otimes\sigma_{B}^{x,y})\}\\
&=\sum_{y}p(y|x)\textrm{Tr}\{\langle y|\Theta_{YB}^{x}|y\rangle\sigma_{B}^{x,y}\} \\
&= \sum_{y}p(y|x)\textrm{Tr}\{G_{B}^{x,y}\sigma_{B}^{x,y}\},
\end{align*}

where $G_{B}^{x,y}\coloneqq\langle y|\Theta_{YB}^{x}|y\rangle$. In an analogous way:
\begin{equation*}
\begin{aligned}
\textrm{Tr}\{\Theta_{YB}(\sigma_{Y}^{x}\otimes\sigma_{B}^{x})\}=\sum_{y}p(y|x)\textrm{Tr}\{G_{B}^{x,y}\sigma_{B}^{x}\}.
\end{aligned}
\end{equation*}
The above derivations lead the test operator to be considered as $\Theta_{YB}=\sum_{y}|y\rangle\langle y|_{Y}\otimes G_{B}^{x,y}$, i.e., the operator classical on $Y$ achieves the same optimal values as any general operator. Next we try to embed the test operator in the $N_{XY^{|\mathcal{L}||\mathcal{K}|}B}^{(\ell,k)}$ as given in (\ref{dn1}) and expanded as in (\ref{test2}).
\begin{figure*}[!t] 
\begin{align}
\nonumber
N_{XY^{|\mathcal{L}||\mathcal{K}|}B}^{(\ell,k)}&=|x\rangle\langle x|_{X}\otimes\mathbbm{1}_{Y}^{(1,1)}\otimes...\otimes \Theta_{YB}^{(\ell,k)}\otimes...\otimes \mathbbm{1}_{Y}^{(|\mathcal{L}|,|\mathcal{K}|)} \\
&= \sum_{y_{11}...y_{lk}...y_{|\mathcal{L}||\mathcal{K}|}}|x\rangle\langle x|_{X}\otimes|y_{11}...y_{\ell k}...y_{|\mathcal{L}||\mathcal{K}|}\rangle\langle y_{11}...y_{\ell k}...y_{|\mathcal{L}||\mathcal{K}|}|\otimes G_{B}^{x,y_{\ell k}}.
\label{test2}
\end{align}
\hrulefill
\end{figure*}
Observe the structure of the POVM given in (\ref{POVMdis}).
\begin{figure*}[!t]
\begin{align}
\nonumber
&\left(\sum_{\ell'}\sum_{k'}N_{XY^{|\mathcal{L}||\mathcal{K}|}B}^{(\ell',k')}\right)^{-\frac{1}{2}}\\
&= \sum_{y_{11}...y_{|\mathcal{L}||\mathcal{K}|}}|x\rangle\langle x|_{X}\otimes|y_{11}...y_{|\mathcal{L}||\mathcal{K}|}\rangle\langle y_{11}...y_{|\mathcal{L}||\mathcal{K}|}| \otimes\left(\sum_{\ell'}\sum_{k'}G_{B}^{x,y_{\ell'k'}}\right)^{-\frac{1}{2}},
\label{POVMdis}
\end{align}
\hrulefill
\end{figure*}
And finally our POVM has the form given in equation (\ref{scdPOVM}).
\begin{figure*}[!t]
\begin{align}
\label{scdPOVM}
\nonumber
P&_{XY^{|\mathcal{L}||\mathcal{K}|}B}^{(\ell,k)}\\ \nonumber
&=\left(\sum_{\ell'}\sum_{k'}N_{XY^{|\mathcal{L}||\mathcal{K}|}B}^{(\ell',k')}\right)^{-\frac{1}{2}}N_{XY^{|\mathcal{L}||\mathcal{K}|}B}^{(\ell,k)}
\left(\sum_{l'}\sum_{k'}N_{XY^{|\mathcal{L}||\mathcal{K}|}B}^{(\ell',k')}\right)^{-\frac{1}{2}}\\
 &=\sum_{y_{11}...y_{|\mathcal{L}||\mathcal{K}|}}|x\rangle\langle x|_{X}\otimes|y_{11}...y_{|\mathcal{L}||\mathcal{K}|}\rangle\langle y_{11}...y_{|\mathcal{L}||\mathcal{K}|}| \otimes \Omega_{B}^{x,y_{\ell k}}, \\ \nonumber
&\text{where}\\ \nonumber
&\Omega_{B}^{x,(\ell,k)}\coloneqq\left(\sum_{\ell'}\sum_{k'}G_{B}^{x,y_{\ell' k'}}\right)^{-\frac{1}{2}}G_{B}^{x,y_{\ell k}} \left(\sum_{\ell'}\sum_{k'}G_{B}^{x,y_{\ell' k'}}\right)^{-\frac{1}{2}}.
\end{align}
\hrulefill
\end{figure*}
To build a POVM on the full space, we add $\Omega_{B}^{0}=\mathbbm{1}_{B}-\sum_{\ell}\sum_{k}\Omega_{B}^{x,(\ell,k)}$ to the set $\{\Omega_{B}^{x,(\ell,k)}\}_{\ell=1,k=1}^{|\mathcal{L}|,|\mathcal{K}|}$.  
By combining (\ref{scdstate}) and (\ref{scdPOVM}), we find that
\begin{align*}
\textrm{Tr}\{(\mathbbm{1}&_{Y^{|\mathcal{L}||\mathcal{K}|}B}-P_{XY^{|\mathcal{L}||\mathcal{K}|}B}^{(\ell,k)})\sigma_{XY^{\otimes|\mathcal{L}||\mathcal{K}|}B}^{m,(\ell,k)}\}\\
&=\sum_{x,y_{11}...y_{|\mathcal{L}||\mathcal{K}|}}p_{X}(x)p(y_{11}|x)...p(y_{|\mathcal{L}||\mathcal{K}|}|x)\\
&\hspace*{2.5cm}\times\textrm{Tr}\{(\mathbbm{1}_{B}-\Omega_{B}^{x,y_{\ell k}})\sigma_{B}^{x,m,y_{\ell k}}\}
\end{align*}
and from (\ref{eqai}), the equality of the average and the individual error probabilities, yields equation (\ref{avlky}).
\begin{figure*}[!t]
\begin{align}
\label{avlky}
\nonumber
\frac{1}{|\mathcal{L}|}\frac{1}{|\mathcal{K}|}\sum_{\ell=1}^{|\mathcal{L}|}\sum_{k=1}^{|\mathcal{K}|}&\textrm{Tr}\{(\mathbbm{1}_{Y^{|\mathcal{L}||\mathcal{K}|}B}-P_{XY^{|\mathcal{L}||\mathcal{K}|}B}^{(\ell,k)})\sigma_{XY^{\otimes|\mathcal{L}||\mathcal{K}|}B}^{m,(\ell,k)}\}\\ \nonumber
&=\frac{1}{|\mathcal{L}|}\frac{1}{|\mathcal{K}|}\sum_{\ell=1}^{|\mathcal{L}|}\sum_{k=1}^{|\mathcal{K}|}\sum_{x,y_{11}...y_{|\mathcal{L}||\mathcal{K}|}}p_{X}(x)p(y_{11}|x)...p(y_{|\mathcal{L}||\mathcal{K}|}|x)\textrm{Tr}\{(\mathbbm{1}_{B}-\Omega_{B}^{x,y_{\ell k}})\sigma_{B}^{x,m,y_{\ell k}}\} \nonumber \\
&=\sum_{x,y_{11}...y_{|\mathcal{L}||\mathcal{K}|}}p_{X}(x)p(y_{11}|x)...p(y_{|\mathcal{L}||\mathcal{K}|}|x)\left(\frac{1}{|\mathcal{L}|}\frac{1}{|\mathcal{K}|}\sum_{\ell=1}^{|\mathcal{L}|}\sum_{k=1}^{|\mathcal{K}|}\textrm{Tr}\{(\mathbbm{1}_{B}-\Omega_{B}^{x,y_{\ell,k}})\sigma_{B}^{x,m,y_{\ell k}}\}\right) \leq \epsilon.
\end{align}
\hrulefill
\end{figure*}

By taking advantage of the POVMs $\{\Omega_{B}^{x,(\ell,k)}\}_{\ell=1,k=1}^{|\mathcal{L}|,|\mathcal{K}|}$, the following measurement channels are defined
\begin{equation}
\mathcal{D'}_{B\rightarrow\hat{L}}^{2}(\omega_{B}) \coloneqq\sum_{\ell=1}^{|\mathcal{L}|}\sum_{k=1}^{|\mathcal{K}|}\textrm{Tr}\{\Omega_{B}^{x,y_{\ell,k}}\omega_{B}\}|\ell\rangle\langle \ell|_{\hat{L}},
\label{mc1}
\end{equation}
\begin{equation}
\mathcal{D'}_{B\rightarrow\hat{L}\hat{K}}^{2}(\omega_{B})\coloneqq\sum_{\ell=1}^{|\mathcal{L}|}\sum_{k=1}^{|\mathcal{K}|}\textrm{Tr}\{\Omega_{B}^{x,y_{\ell,k}}\omega_{B}\}|\ell\rangle\langle \ell|_{\hat{L}}\otimes |k\rangle\langle k|_{\hat{K}},
\end{equation}
where $\omega_{B}$ is a general quantum state and $\textrm{Tr}_{\hat{K}}\circ\mathcal{D'}_{B\rightarrow\hat{L}\hat{K}}^{2}=\mathcal{D'}_{B\rightarrow\hat{L}}^{2}$. Note that in (\ref{mc1}) the probability of getting a particular $\ell$ equals $\sum_{k=1}^{|\mathcal{K}|}\textrm{Tr}\{\Omega_{B}^{x,y_{\ell k}}\omega_{B}\}$. By direct calculations, it can be seen that:
\begin{align*}
\textrm{Tr}&\{(\mathbbm{1}_{B}-\Omega_{B}^{x,y_{\ell,k}})\sigma_{B}^{x,m,y_{\ell,k}}\}\\
&=\frac{1}{2}\left\|\mathcal{D'}_{B\rightarrow\hat{L}\hat{K}}^{2}(\sigma_{B}^{x,m,y_{\ell,k}})-|\ell\rangle\langle \ell|_{\hat{L}}\otimes |k\rangle\langle k|_{\hat{K}}\right\|_{1},
\end{align*}
averaging over $\ell$, $k$ and $(x,y_{1,1}...y_{|\mathcal{L}|,|\mathcal{K}|})$ and using (\ref{avlky}), we get equation (\ref{figure2}).
\begin{figure*}[!t]
\begin{equation}
\label{figure2}
\sum_{x,y_{11}...y_{|\mathcal{L}||\mathcal{K}|}}p_{X}(x)p(y_{11}|x)...p(y_{|\mathcal{L}||\mathcal{K}|}|x)\left[\frac{1}{|\mathcal{L}|}\frac{1}{|\mathcal{K}|}\sum_{\ell}\sum_{k}\frac{1}{2}\left\|\mathcal{D'}_{B\rightarrow\hat{L}\hat{K}}^{2}(\sigma_{B}^{x,m,y_{\ell k}})-|\ell\rangle\langle \ell|_{\hat{L}}\otimes |k\rangle\langle k|_{\hat{K}}\right\|_{1}\right]\leq \epsilon.
\end{equation}
\hrulefill
\end{figure*}
In equation (\ref{figure2}), if we take the average over $k$ inside the trace distance and trace out $\hat{K}$ system, by the convexity and monotonicity of the trace distance, we obtain the equations in (\ref{avklepsilon}).
\begin{figure*}[!t]
\begin{align}
\label{avklepsilon}
\nonumber
\sum_{x,y_{11},...,y_{|\mathcal{L}||\mathcal{K}|}}&p_{X}(x)p(y_{11}|x)...p(y_{|\mathcal{L}||\mathcal{K}|}|x)\left[\frac{1}{|\mathcal{L}|}\sum_{\ell=1}^{|\mathcal{L}|}\frac{1}{2}\left\|(\textrm{Tr}_{\hat{K}}\circ\mathcal{D'}_{B\rightarrow\hat{L}\hat{K}}^{2})\left(\frac{1}{|\mathcal{K}|}\sum_{k=1}^{|\mathcal{K}|}\sigma_{B}^{x,m,y_{\ell k}}\right)-|\ell\rangle\langle \ell|_{\hat{L}}\right\|_{1}\right] \\
&=\sum_{x,y_{11},...,y_{|\mathcal{L}||\mathcal{K}|}}p_{X}(x)p(y_{11}|x)...p(y_{|\mathcal{L}||\mathcal{K}|}|x)\left[\frac{1}{|\mathcal{L}|}\sum_{\ell=1}^{|\mathcal{L}|}\frac{1}{2}\left\|\mathcal{D'}^{2}_{B\rightarrow\hat{L}}\left(\frac{1}{|\mathcal{K}|}\sum_{k=1}^{|\mathcal{K}|}\sigma_{B}^{x,m,y_{\ell k}}\right)-|\ell\rangle\langle \ell|_{\hat{L}}\right\|_{1}\right]\leq \epsilon.
\end{align}
\hrulefill
\end{figure*}

Considering the POVM $\{\Omega_{B}^{x,y_{\ell k}}\}_{\ell=1,k=1}^{|\mathcal{L}|,|\mathcal{K}|}$, the probability of getting $\ell'$ conditioned on the fact that $(\ell,k)$ was sent is equal to $\sum_{k'=1}^{K}\textrm{Tr}\{\Omega_{B}^{x,y_{\ell' k'}}\sigma_{B}^{x,y_{\ell k}}\}$ and it is clear from the uniformity of the local key that the probability of getting $l'$ given that $l$ was sent, equals $Pr(l'|l)=\frac{1}{|\mathcal{K}|}\sum_{k=1}^{|\mathcal{K}|}\sum_{k'=1}^{|\mathcal{K}|}\textrm{Tr}\{\Omega_{B}^{x,y_{\ell' k'}}\sigma_{BE}^{x,m,y_{\ell k}}\}=\frac{1}{|\mathcal{K}|}\sum_{k=1}^{|\mathcal{K}|}\sum_{k'=1}^{|\mathcal{K}|}\textrm{Tr}\{\Omega_{B}^{x,y_{\ell' k'}}\sigma_{B}^{x,m,y_{\ell k}}\}$. Note that evidently $\sum_{\ell'=1}^{|\mathcal{L}|}Pr(\ell'|\ell)=1$.
If the trace above was only applied to the $B$ system, we would have :
\begin{equation*}
\frac{1}{|\mathcal{K}|}\sum_{k=1}^{|\mathcal{K}|}\sum_{k'=1}^{|\mathcal{K}|}\textrm{Tr}_{B}\{\Omega_{B}^{x,y_{\ell',k'}}\sigma_{BE}^{x,m,y_{\ell,k}}\}=Pr(\ell'|\ell)u_{E}^{\ell',\ell},
\end{equation*}
where 
\begin{align*}
u_{E}^{\ell',\ell}\coloneqq\frac{\frac{1}{|\mathcal{K}|}\sum_{k=1}^{|\mathcal{K}|}\sum_{k'=1}^{|\mathcal{K}|}\textrm{Tr}_{B}\{\Omega_{B}^{x,y_{\ell',k'}}\sigma_{BE}^{x,m,y_{\ell,k}}\}}{Pr(\ell'|\ell)}.
\end{align*}
And by summing up over all $\ell'$ we get: (see that $\sum_{k'=1}^{|\mathcal{K}|}\sum_{\ell'=1}^{|\mathcal{L}|}\textrm{Tr}_{B}\{\Omega_{B}^{x,y_{\ell',k'}}\sigma_{BE}^{x,m,y_{\ell,k}}\}=\sigma_{E}^{x,m,y_{\ell,k}}$ for a given pair $(\ell,k)$)
\begin{equation*}
\sum_{\ell'=1}^{|\mathcal{L}|}Pr(\ell'|\ell)u_{E}^{\ell',\ell}=\frac{1}{|\mathcal{K}|}\sum_{k=1}^{|\mathcal{K}|}\sigma_{E}^{x,m,y_{\ell,k}}.
\end{equation*}
Hence, the following equation follows:
\begin{equation*}
\mathcal{D'}_{B\rightarrow\hat{L}}^{2}\left(\frac{1}{|\mathcal{K}|}\sum_{k=1}^{|\mathcal{K}|}\sigma_{BE}^{x,m,y_{\ell,k}}\right)=\sum_{\ell'=1}^{|\mathcal{L}|}Pr(\ell'|\ell)|\ell'\rangle\langle \ell'|_{\hat{L}}\otimes u_{E}^{\ell',\ell},
\end{equation*}
and by tracing out Eve's system:
\begin{equation*}
\mathcal{D'}_{B\rightarrow\hat{L}}^{2}\left(\frac{1}{|\mathcal{K}|}\sum_{k=1}^{|\mathcal{K}|}\sigma_{B}^{x,m,y_{\ell,k}}\right)=\sum_{\ell'=1}^{|\mathcal{L}|}Pr(\ell'|\ell)|\ell'\rangle\langle \ell'|_{\hat{L}}.
\end{equation*}
We move forward with the chain of inequalities ending up in (\ref{chain1}),
\begin{figure*}[!t]
\begin{align}
\label{chain1}
\nonumber
\frac{1}{|\mathcal{L}|}\sum_{l=1}^{|\mathcal{L}|}\frac{1}{2}&\left\|\mathcal{D'}_{B\rightarrow\hat{L}}^{2}\left(\frac{1}{|\mathcal{K}|}\sum_{k=1}^{|\mathcal{K}|}\sigma_{BE}^{x,m,y_{\ell,k}}\right)-|\ell\rangle\langle\ell|_{\hat{L}}\otimes \frac{1}{|\mathcal{K}|}\sum_{k=1}^{|\mathcal{K}|}\sigma_{E}^{x,m,y_{\ell,k}}\right\|_{1} \\ \nonumber
&=\frac{1}{|\mathcal{L}|}\sum_{\ell=1}^{|\mathcal{L}|}\frac{1}{2}\left\|\sum_{\ell'=1}^{|\mathcal{L}|}Pr(\ell'|\ell)|\ell'\rangle\langle\ell'|_{\hat{L}} \otimes u_{E}^{\ell',\ell}-|\ell\rangle\langle\ell |_{\hat{L}}\otimes \sum_{\ell'=1}^{|\mathcal{L}|}Pr(\ell'|\ell)u_{E}^{\ell',\ell}\right\|_{1}  \\ \nonumber
&\leq\frac{1}{|\mathcal{L}|}\sum_{l=1}^{|\mathcal{L}|}\sum_{\ell'=1}^{|\mathcal{L}|}Pr(\ell'|\ell)\left[\frac{1}{2}\||\ell'\rangle\langle\ell'|_{\hat{L}}\otimes u_{E}^{\ell',\ell}-|\ell\rangle\langle\ell|_{\hat{L}}\otimes u_{E}^{\ell',\ell}\|_{1}\right]=\frac{1}{|\mathcal{L}|}\sum_{\ell=1}^{|\mathcal{L}|}\sum_{\ell'=1}^{|\mathcal{L}|}Pr(\ell'|\ell)\left[\frac{1}{2}\||\ell'\rangle\langle\ell'|_{\hat{L}}-|\ell\rangle\langle\ell|_{\hat{L}}\|_{1}\right] \\
&=\frac{1}{|\mathcal{L}|}\sum_{\ell=1}^{|\mathcal{L}|}\sum_{\ell'\ne\ell}Pr(\ell'|\ell)=\frac{1}{|\mathcal{L}|}\sum_{\ell=1}^{|\mathcal{L}|}\frac{1}{2}\left\|\mathcal{D'}_{B\rightarrow\hat{L}}^{2}\left(\frac{1}{|\mathcal{K}|}\sum_{k=1}^{|\mathcal{K}|}\sigma_{B}^{x,m,y_{\ell,k}}\right)-|\ell\rangle\langle \ell |_{\hat{L}}\right\|_{1},
\end{align}
\hrulefill
\end{figure*}
where the first inequality follows from the convexity of the trace distance and the second equality emerges because of  the invariance of the trace distance with respect to tensor-product states. This result together with (\ref{avklepsilon}) leads to the inequality given by (\ref{Bob-et}).
\begin{figure*}[!t]
\begin{align}
\label{Bob-et}
\sum_{x,y_{1,1},...,y_{|\mathcal{L}|,|\mathcal{K}|}}&p_{X}(x)p(y_{1,1}|x)...p(y_{|\mathcal{L}|,|\mathcal{K}|}|x)\left[\frac{1}{|\mathcal{L}|}\sum_{\ell=1}^{|\mathcal{L}|}\frac{1}{2}\left\|\mathcal{D'}_{B\rightarrow\hat{L}}^{2}\left(\frac{1}{|\mathcal{K}|}\sum_{k=1}^{|\mathcal{K}|}\sigma_{BE}^{x,m,y_{\ell,k}}\right)-|\ell\rangle\langle\ell|_{\hat{L}}\otimes \frac{1}{|\mathcal{K}|}\sum_{k=1}^{|\mathcal{K}|}\sigma_{E}^{x,m,y_{\ell,k}}\right\|_{1}\right]\leq \epsilon.
\end{align}
\hrulefill
\end{figure*}
This is equivalent to the criterion dealing with Bob's error in detecting the private message. We continue by expanding Eve's security condition as given in (\ref{Eve-st}),
\begin{figure*}[!t]
\begin{align}
\label{Eve-st}
\nonumber
&\frac{1}{2}\left\|\sigma_{XY^{\otimes|\mathcal{L}||\mathcal{K}|}E}^{m,\ell}-\sum_{x}p_{X}(x)|x\rangle\langle x|_{X}\otimes\sigma_{Y^{\otimes|\mathcal{L}||\mathcal{K}|}}^{x,m}\otimes\tilde{\sigma}_{E}^{x,m}\right\|_{1} \nonumber \\
&=\frac{1}{2}\left\|\frac{1}{|\mathcal{K}|}\sum_{k=1}^{|\mathcal{K}|}\sum_{x,y_{11}...y_{|\mathcal{L}||\mathcal{K}|}}p_{X}(x)p_{Y|X}(y_{11}|x)...p_{Y|X}(y_{|\mathcal{L}||\mathcal{K}|}|x)|x\rangle\langle x|_{X}\otimes|y_{11}...y_{|\mathcal{L}||\mathcal{K}|}\rangle\langle y_{11}...y_{|\mathcal{L}||\mathcal{K}|}|_{Y^{|\mathcal{L}||\mathcal{K}|}}\otimes(\sigma_{E}^{x,m,y_{\ell k}}-\tilde{\sigma}_{E}^{x,m})\right\|_{1} \nonumber \\
&=\frac{1}{2}\left\|\sum_{x,y_{11}...y_{\ell k}}p_{X}(x)p_{Y|X}(y_{11}|x)...p_{Y|X}(y_{|\mathcal{L}|,|\mathcal{K}|}|x)|x\rangle\langle x|_{X}\otimes|y_{11}...y_{|\mathcal{L}||\mathcal{K}|}\rangle\langle y_{11}...y_{|\mathcal{L}|,|\mathcal{K}|}|_{Y^{|\mathcal{L}||\mathcal{K}|}}\otimes\left(\frac{1}{|\mathcal{K}|}\sum_{k=1}^{|\mathcal{K}|}\sigma_{E}^{x,m,y_{\ell,k}}-\tilde{\sigma}_{E}^{x,m}\right)\right\|_{1} \nonumber \\
&=\sum_{x,y_{11}...y_{|\mathcal{L}||\mathcal{K}|}}p_{X}(x)p_{Y|X}(y_{11}|x)...p_{Y|X}(y_{|\mathcal{L}||\mathcal{K}|}|x)\left[\frac{1}{2}\left\|\frac{1}{|\mathcal{K}|}\sum_{k=1}^{|\mathcal{K}|}\sigma_{E}^{x,m,y_{\ell k}}-\tilde{\sigma}_{E}^{x,m}\right\|_{1}\right]\nonumber \\
&=\sum_{x,y_{11}...y_{|\mathcal{L}||\mathcal{K}|}}p_{X}(x)p_{Y|X}(y_{11}|x)...p_{Y|X}(y_{|\mathcal{L}||\mathcal{K}|}|x)\left[\frac{1}{2}\left\||\ell\rangle\langle\ell|_{L'}\otimes\frac{1}{|\mathcal{K}|}\sum_{k=1}^{|\mathcal{K}|}\sigma_{E}^{x,m,y_{\ell k}}-|\ell\rangle\langle\ell|_{L'}\otimes\tilde{\sigma}_{E}^{x,m}\right\|_{1}\right]\leq\sqrt{\epsilon'},
\end{align}
\hrulefill
\end{figure*}
where the last equality comes about by using the invariance of trace distance with respect to tensor-product states. 

We deal with two important expressions in (\ref{Bob-et}) and (\ref{Eve-st}), the former is Bob's error in detecting the private message and the later is the security of Eve. Now it is time to unify two criteria into the so-called privacy error. To this end, let's consider (\ref{Bob-et}) and (\ref{Eve-st}) together with their imposed bounds on the cardinalities of $|\mathcal{L}|$ and $|\mathcal{K}|$. We employ triangle inequality for the trace distance to merge them into the privacy error as given in (\ref{p-Eve1}) (remember that in the assisted code, there is no difference between average and individual error probabilities).
\begin{figure*}[!t]
\begin{align}
\sum_{x,y_{1,1},...,y_{|\mathcal{L}|,|\mathcal{K}|}}p_{X}(x)p_{Y|X}(y_{1,1}|x)...p_{Y|X}(y_{|\mathcal{L}|,|\mathcal{K}|}|x)
\left[\frac{1}{2}\left\|\mathcal{D'}_{B\rightarrow\hat{L}}^{2}\left(\frac{1}{|\mathcal{K}|}\sum_{k=1}^{|\mathcal{K}|}\sigma_{BE}^{x,m,y_{\ell,k}}\right)-|\ell\rangle\langle\ell|_{\hat{L}}\otimes\tilde{\sigma}_{E}^{x,m}\right\|_{1}\right]\leq\epsilon+\sqrt{\epsilon'},
\label{p-Eve1}
\end{align}
\hrulefill
\end{figure*}
This immediately implies the privacy criterion given in (\ref{p-Eve}) in the sense that if this holds, the single criterion in (\ref{p-Eve}) also holds.  

We are now done with the assisted code. As we proceed to derandomize the code, it will be clear that the procedure employed to unify two error criteria is helpful. Before we proceed to derandomize the code, we would like to consider two extra error terms.  The error probability of the second decoder depends on the error probability of the first decoder in two directions, first, the second decoder is fed a state close to the actual received state and second, the second decoder applies a quantum instrument depending on the estimate of the transmitted message $m$. This can, without losing the generality, be written as follows:
\begin{align*}
\mathcal{D}^{2}_{\hat{M}BY\rightarrow\hat{L}}\coloneqq\sum_{m}|m\rangle\langle m|_{M'}\otimes\mathcal{D}^{2,m}_{BY\rightarrow\hat{L}}.
\end{align*}
 In the following we show how this fact contributes to the error probability. First one is the difference between the received state and the disturbed state being fed into the second decoder. Since the probability of error of the first decoder is at most $\epsilon$, we know from gentle measurement lemma that:
\begin{align*}
\left\|\sigma_{XY^{\otimes|\mathcal{L}||\mathcal{K}|}B}^{m,(\ell,k)}-\rho_{XY^{\otimes|\mathcal{L}||\mathcal{K}|}B}^{m,(\ell,k)}\right\|_{1}\leq 2\sqrt{\epsilon},
\end{align*}
and for the second term we have the chain of inequalities given by (\ref{errr});
\begin{figure*}[!t]
\begin{align}
\label{errr}
\nonumber
&\left\|(\sum_{m}|m\rangle\langle m|_{M'}\otimes\mathcal{D}^{2,m}_{BY\rightarrow\hat{L}})(\mathcal{D}^{1}_{BX\rightarrow \hat{M}B}(\rho_{X^{\otimes|\mathcal{M}|}B}^{m,(\ell,k)}))-(|m\rangle\langle m|_{M'}\otimes\mathcal{D}^{2,m}_{BY\rightarrow\hat{L}})(\mathcal{D}^{1}_{BX\rightarrow \hat{M}B}(\rho_{X^{\otimes|\mathcal{M}|}B}^{m,(\ell,k)}))\right\|_{1} \nonumber \\
&\qquad\qquad=\left\|\sum_{m'\neq m}|m'\rangle\langle m'|_{M'}\otimes\mathcal{D}^{2,m'}_{BY\rightarrow\hat{L}}(\mathcal{D}^{1,m'}_{BX\rightarrow B}(\rho_{X^{\otimes|\mathcal{M}|}B}^{m,(\ell,k)}))\right\|_{1} \nonumber \\
&\qquad\qquad\leq \sum_{m'\neq m}\left\||m'\rangle\langle m'|_{M'}\otimes\mathcal{D}^{2,m'}_{BY\rightarrow\hat{L}}(\mathcal{D}^{1,m'}_{BX\rightarrow B}(\rho_{X^{\otimes|\mathcal{M}|}B}^{m,(\ell,k)}))\right\|_{1} \nonumber \\ 
&\qquad\qquad\leq \sum_{m'\neq m}\left\|\mathcal{D}^{1,m'}_{BX\rightarrow B}(\rho_{X^{\otimes|\mathcal{M}|}B}^{m,(\ell,k)})\right\|_{1}\leq\epsilon.
\end{align}
\hrulefill
\end{figure*}
where the equality follows from the the observation in (\ref{fstd}), the first and second inequalities follow the convexity and monotonicity of trace distance, respectively. 
Adding these two terms to (\ref{p-Eve1}) will result in $P_{priv}\leq 2(\epsilon+\sqrt{\epsilon})+\sqrt{\epsilon'}$.
\subsubsection{Derandomization}
We can now fix the classical registers and obtain a protocol without shared randomness, i.e., derandomize the code. The derandomization is a standard technique and its mathematical details are given in the appendix.  
\section{Converse}
In this section we give upper bounds for the capacity region $\mathcal{C}^{\epsilon,\epsilon'}(\mathcal{N})$.
\begin{IEEEproof}[Proof of theorem (\ref{converse})]
Two messages $m\in\mathcal{M}$ and $\ell\in\mathcal{L}$ are sent through the channel $\mathcal{N}_{A\rightarrow BE}$ and their estimates are $\hat{M}$ and $\hat{L}$, respectively.
From definition (\ref{maind}), an $(\epsilon,\epsilon')$-code satisfies $Pr(M\neq\hat{M})\leq\epsilon$. A hypothesis testing problem can be associated to the problem of detecting $m$ leading to an expression for the error probability of the public message. To see how it works out, consider a binary hypothesis testing problem in which null and alternative hypothesis are 
\begin{align*}
\rho_{MM'}&=\frac{1}{|\mathcal{M}|}\sum_{m=1}^{|\mathcal{M}|}|m\rangle\langle m|_{M}\otimes|m\rangle\langle m|_{M'}\quad\text{and} \\
\rho_{M}\otimes\rho_{M'}&=\frac{1}{|\mathcal{M}|}\sum_{m=1}^{|\mathcal{M}|}|m\rangle\langle m|_{M}\otimes\frac{1}{|\mathcal{M}|}\sum_{m=1} ^{|\mathcal{M}|}|m\rangle\langle m|_{M'},
\end{align*}
respectively. It is easily seen that type I error, i.e., deciding in favor of $\rho_{M}\otimes\rho_{M'}$ while the true state was $\rho_{MM'}$, is exactly equal to the error probability $Pr(M\neq\hat{M})$ which is less than or equal to $\epsilon$ by assumption. On the other hand, type II error, deciding $\rho_{MM'}$ on $\rho_{M}\otimes\rho_{M'}$, equals $\frac{1}{|\mathcal{M}|}$ (the distribution over message set is uniform). Then from the definition of the hypothesis testing mutual information, we have the following:
\begin{align*}
r\leq I_{H}^{\epsilon}(M;M^{\prime})_{\rho}.
\end{align*}
where $r=\log|\mathcal{M}|$ is the rate of the public message. Furthermore, from the quantum DPI, we have:
\begin{align*}
I_{H}^{\epsilon}(M;M^{\prime})_{\rho}\leq I_{H}^{\epsilon}(M;B)_{\rho}
\end{align*} 
Finally, using the injectivity of the encoder, we define a random variable $X$ whose distribution is built by projecting the distribution of $M$ on its image on $X$ and zero otherwise. Setting $X=M$, we get the following:
\begin{align*}
r\leq I_{H}^{\epsilon}(X;B)_{\rho}.
\end{align*}

In regards to the private rate $R=\log|\mathcal{L}|$, consider the following chain of inequalities:
 \begin{align*}
\epsilon&\geq Pr\{(M,L)\neq (\hat{M},\hat{L})\} \\
&=\sum_{m,\ell}p(m)p(\ell)\sum_{(\hat{m},\hat{\ell})\neq(m,\ell)}p(\hat{m},\hat{\ell}|m,\ell) \\
&\geq\sum_{m,\ell}p(m)p(\ell)\sum_{\hat{\ell}\neq\ell}p(\hat{\ell}|m,\ell) \\
&=\sum_{m}p(m)Pr(\hat{L}\neq L|M=m),
\end{align*}
where the first line is due to the assumption. From Markov's inequality, we know that with probability at least $1-\sqrt{\epsilon}$, the following holds for a randomly generated $m\in\mathcal{M}$:
\begin{align*}
Pr(\hat{L}\neq L|M=m)\leq\sqrt{\epsilon}.
\end{align*}
Then following the same strategy as for the public rate, we consider a binary hypothesis testing problem distinguishing between $\rho_{L\hat{L}}^{m}$ and $\rho_{L}^{m}\otimes\rho_{L}^{m}$ conditioned on previously specified $m$, we will have:
\begin{align*}
R\leq I_{H}^{\sqrt{\epsilon}}(L;\hat{L}|M=m)_{\rho}.
\end{align*}
Then, to get $I_{H}^{\sqrt{\epsilon}}(L;\hat{L}|M)_{\rho}$, according to Definition \ref{chtma}, we can optimize the expression with respect to $\rho_{M}'$ where $P(\rho_{M},\rho_{M}')\leq\sqrt{\epsilon}$.
Then from the monotonicity of the hypothesis-testing relative entropy applied to $\hat{L}$ system, we have:
\begin{align*}
I_{H}^{\sqrt{\epsilon}}(L;\hat{L}|M)_{\rho}\leq I_{H}^{\sqrt{\epsilon}}(L;B|M)_{\rho}.
\end{align*}
By the same argument that we defined $X\coloneqq M$, we also define $Y\coloneqq L$ and so the following results:
\begin{align}
\label{p1}
R\leq I_{H}^{\sqrt{\epsilon}}(Y;B|X)_{\rho}.
\end{align}
On the other hand, from the secrecy condition (\ref{d2}), we know that for every $m$, the following is true:
\begin{align*}
\frac{1}{2}\|\rho_{LE}^{m}-\rho_{L}\otimes\tilde{\rho}_{E}^{m}\|_{1}\leq\epsilon',
\end{align*}
and from the relation between the purified distance and the trace distance it holds that:
\begin{align*}
P(\rho_{LE}^{m},\rho_{L}\otimes\tilde{\rho}_{E}^{m})\leq\sqrt{2\epsilon'}.
\end{align*}
From the definition of the smooth max-relative entropy we see that $D_{max}^{\sqrt{2\epsilon'}}(\rho_{LE}^{m},\rho_{L}\otimes\tilde{\rho}_{E}^{m})=0$. And by considering the optimization in Definition \ref{casmma} over $\rho_{M}^{\prime}$ such that $P(\rho_{M}^{\prime},\rho_{M})\leq\sqrt{\epsilon'}$, we have $I_{max}^{\sqrt{2\epsilon'}}(L;E|M)=0$.
Setting $M\coloneqq X$ and $L\coloneqq Y$ as before and plugging into (\ref{p1}), the following bound on the private rate holds:
\begin{align}
R\leq I_{H}^{\sqrt{\epsilon}}(Y;B|X)-I_{max}^{\sqrt{2\epsilon'}}(Y;E|X).
\end{align}

\end{IEEEproof}

\section{Asymptotic Analysis}
We evaluate our rate region in the asymptotic limit of many uses of a memoryless channel. The capacity theorem for simultaneous transmission of classical and quantum information was proved by Devetak and Shor \cite{DevShor}. In this section, we recover their result from our theorems. We define the rate region of the simultaneous transmission of the classical and quantum information as follows: 
\begin{align*}
\mathcal{C}_{\infty}(\mathcal{N})\coloneqq\lim_{\epsilon,\epsilon'\rightarrow 0}\lim_{n\rightarrow\infty}\frac{1}{n}\mathcal{C}^{\epsilon,\epsilon'}(\mathcal{N}^{\otimes n}).
\end{align*}

Let $\mathcal{C}(\mathcal{N})$ be the set of rate pairs $(r',R')$ such that
\begin{align*}
r'\leq& I(X;B)_{\rho},\\
R'\leq&  I\left(Y;B|X\right)_{\rho}-I\left(Y;E|X\right)_{\rho}
\end{align*}
where all the entropic quantities are computed over all $\rho_{XYBE}\coloneqq\sum_{x,y}p(x,y)\ketbra{x}\otimes\ketbra{y}\otimes\mathcal{N}_{A\rightarrow BE}(\rho_{A}^{x,y})$ arising from the channel. Then the capacity region $\mathcal{C}_{\infty}(\mathcal{N})$ is the (normalized) union over $\ell$ uses of the channel $\mathcal{N}$ as below:
\begin{align}
\label{r11}
\mathcal{C}_{\infty}(\mathcal{N})=\bigcup_{\ell=1}^{\infty}\frac{1}{\ell}\mathcal{C}(\mathcal{N}^{\otimes \ell}).
\end{align}
In the rest of this section, our aim is to prove the capacity region above. Before doing so, we slightly modify the expression for the private rate in Theorem \ref{achievability tm} by using Fact \ref{fact4}. Note that Fact \ref{fact4} deals with unconditional expressions, however, conditional expressions are trivial noting their definitions. Therefore, the following holds:
\begin{align*}
\tilde{I}_{max}^{\sqrt{\epsilon'}-\delta'}(Y;E|X)\leq I_{max}^{\sqrt{\epsilon'}-\delta'-\gamma}(Y;E&|X)\\
&+\log_{2}\left(\frac{3}{\gamma^{2}} \right),
\end{align*}
where $\gamma\in(0, \sqrt{\epsilon'}-\delta')$. And so the achievability of the private rate appears as follows:
\begin{align*}
R&\geq I_{H}^{\epsilon-\delta}\left(Y;B|X\right)-I_{max}^{\sqrt{\epsilon'}-\delta'-\gamma}(Y;E|X)\\
&\hspace*{2cm}-\log_{2}(\frac{4\epsilon}{\delta^{2}})-2\log_{2}(\frac{1}{\delta'})-\log_{2}\left(\frac{3}{\gamma^{2}} \right).
\end{align*}

Like all capacity theorems, the proof of the aforementioned capacity region is accomplished in two steps, direct part that we show all such rates are achievable, i.e., the right-hand side of the equation (\ref{r11}) is contained ($\subseteq$) inside $\mathcal{C}_{\infty}(\mathcal{N})$ and the converse part that goes in the opposite direction saying that those rates cannot be exceeded, i.e., $\mathcal{C}_{\infty}(\mathcal{N})$ is contained inside the union on the right-hand side of (\ref{r11}).  

  For the direct part, we use our one-shot lower bounds on the capacity region and apply quantum AEP for the (conditional) smooth hypothesis testing- and max-mutual information. From Theorem \ref{achievability tm}, for $m$ uses of the channel $\mathcal{N}$ (or as one may like to think of it, one use of the superchannel $\mathcal{N}^{\otimes m}$),  the following lower bound on the capacity region $\mathcal{C}^{\epsilon,\epsilon'}(\mathcal{N}^{\otimes m})$ can be seen:
  \begin{align*}
\bigcup_{\ell=1}^{m}\mathcal{C}_{a}(\mathcal{N}^{\otimes \ell})\subseteq\mathcal{C}^{3\epsilon+2\sqrt{\epsilon}+\sqrt{\epsilon'},2(\epsilon+\sqrt{\epsilon})+\sqrt{\epsilon'}}(\mathcal{N}^{\otimes m}),
\end{align*}
where $\mathcal{C}_{a}(\mathcal{N}^{\otimes \ell})$ is the set of all twins $(r',R')$ satisfying:
\begin{align*}
r'&\leq I_{H}^{\epsilon-\delta}\left(X^{\ell};B^{\otimes \ell}\right)-\log_{2}(\frac{4\epsilon}{\delta^{2}}),\\
R'&\leq I_{H}^{\epsilon-\delta}\left(Y^{\ell};B^{\otimes \ell}|X^{\ell}\right)-I_{max}^{\sqrt{\epsilon'}-\delta'-\gamma}(Y^{\ell};E^{\otimes \ell}|X^{\ell})\\
&\hspace*{2cm}-\log_{2}(\frac{4\epsilon}{\delta^{2}})-2\log_{2}(\frac{1}{\delta'})-\log_{2}\left(\frac{3}{\gamma^{2}} \right).\\
\end{align*}
Since the region above is basically a lower bound on the capacity region, we are free to assume that the sequences of the random variables are generated in an $i.i.d.$ fashion according to the corresponding distributions. This empowers us to make use of quantum AEP as described below. From Fact \ref{factdos} we have 
\begin{align*}
\lim_{\epsilon\rightarrow 0}\lim_{m\rightarrow\infty}\frac{1}{m} I_{H}^{\epsilon-\delta}\left(X^{m};B^{\otimes m}\right)=I(X;B).
\end{align*}
Likewise, applying Lemma \ref{l1} and Lemma \ref{l2} give rise respectively to the following identities:
\begin{align*}
\lim_{\epsilon\rightarrow 0}\lim_{m\rightarrow\infty}\frac{1}{m} I_{H}^{\epsilon-\delta}\left(Y^{m};B^{\otimes m}|X^{m}\right)=I(Y;B|X),\\
\lim_{\epsilon'\rightarrow 0}\lim_{m\rightarrow\infty}\frac{1}{m} I_{max}^{\sqrt{\epsilon'}-\delta'-\gamma}(Y^{m};E^{\otimes m}|X^{m})=I(Y;E|X).
\end{align*}
Plugging back into the respective equations, we obtain 
\begin{align*}
\mathcal{C}(\mathcal{N})\subseteq\lim_{\epsilon,\epsilon'\rightarrow 0}\lim_{m\rightarrow\infty}\frac{1}{m}\mathcal{C}^{\epsilon,\epsilon'}(\mathcal{N}^{\otimes m}),
\end{align*}
where $\mathcal{C}(\mathcal{N})$, as defined before, consists of rate pairs $(r',R')$ satisfying
\begin{align*}
r'\leq& I(X;B)_{\rho},\\
R'\leq&I(Y;B|X)_{\rho}-I(Y;E|X)_{\rho}.
\end{align*}
Last step of the direct part is to consider a superchannel $\mathcal{N}^{\otimes \ell}$ ($\ell$ independent uses of the channel $\mathcal{N}$) and let $n=m\ell$ and repeat the above argument, i.e., use the superchannel $m$ times. Finally by letting $n\rightarrow\infty$ and evaluating the union of the regions, we obtain the desired result.

To prove the converse, we consider our upper bounds given in Theorem \ref{converse} in the case of $n$ uses of the channel $\mathcal{N}$ and we have:
\begin{align*}
\mathcal{C}^{\epsilon,\epsilon'}(\mathcal{N}^{\otimes n})\subseteq\bigcup_{n=1}^{\infty}\mathcal{C}_{c}(\mathcal{N}^{\otimes n})
\end{align*}
where $\mathcal{C}_{c}(\mathcal{N}^{\otimes n})$ includes all ordered twins $(r',R')$ satisfying 
\begin{align}
\label{rr}
r'&\leq I_{H}^{\epsilon}\left(X^{n};B^{\otimes n}\right),\\
\label{RR}
R'&\leq I_{H}^{\epsilon}(Y^{n};B^{\otimes n}|X^{n})-I_{max}^{\sqrt{2\epsilon'}}\left(E^{\otimes n};Y^{n}|X^{n}\right).
\end{align}
To upper bound right-hand side of (\ref{rr}) we apply Fact \ref{rbqh}. The first term on the right-hand side of (\ref{RR}) can be upper bounded by making use of lemma (\ref{lh}) and for the second term, we use Lemma \ref{lm} replacing $|\mathcal{H}_{A}|$ with $|\mathcal{Y}|^{n}$. The inequalities are as follows:
\begin{align*}
r'&\leq \frac{1}{1-\epsilon}\left( I(X^{n};B^{\otimes n})+h_{b}(\epsilon)\right),\\
R'&\leq \frac{1}{1-\epsilon}\left(I(Y^{n};B^{\otimes n}|X^{n})+h_{b}(\epsilon)\right)-I\left(E^{\otimes n};Y^{n}|X^{n}\right) \\
&+2n\sqrt{2\epsilon'}\log|\mathcal{Y}|+2(1+\sqrt{2\epsilon'})h_{b}(\frac{\sqrt{2\epsilon'}}{1+\sqrt{2\epsilon'}}).
\end{align*}
Multiplying by $\frac{1}{n}$ and taking the limits $n\rightarrow\infty$ and $\epsilon,\epsilon'\rightarrow 0$, (changing $n$ with $\ell$) the desired result is achieved.

\subsection{private information to coherent information}
Here we argue that the private rate that has been given in terms of the difference between two mutual-information like quantities, is in principle, the coherent information appearing in \cite{DevShor}. To see how this plays out, consider an ensemble of quantum states $\mathcal{E}=\{p_{X}(x), |\phi^{x}\rangle_{RA}\}_{x\in \mathcal{X}}$ where $X$ is a random variable with alphabet $\mathcal{X}$ and distribution $p_{X}(x)$ and $A$ and $R$ are quantum systems such that $R$ plays the role of a reference system. Assuming an auxiliary classical system $\sigma_{X}=\sum_{x}p_{X}(x)|x\rangle\langle x|_{X}$, the following state can be associated to the ensemble:
\begin{align}
\label{icoh}
\sigma_{XRA}=\sum_{x}p_{X}(x)|x\rangle\langle x|_{X}\otimes |\phi^{x}\rangle\langle\phi^{x}|_{RA}.
\end{align}
If channel $\mathcal{N}_{A\rightarrow BE}$ acts on this state, we get the following \textit{coherent} state:
\begin{align*}
\mathcal{N}_{A\rightarrow BE}(\sigma_{XRA})=\sum_{x}p_{X}(x)|x\rangle\langle x|_{X}\otimes |\phi^{x}\rangle\langle\phi^{x}|_{RBE},
\end{align*}
and the conditional coherent information on it, is evaluated as follows: 
\begin{align*}
I(R\rangle BX)\coloneqq -H(R|BX)&=H(B|X)-H(RB|X) \\
&\stackrel{(a)}{=}H(B|X)-H(E|X),
\end{align*}
where $(a)$ follows from the fact that the state $|\phi^{x}\rangle\langle\phi^{x}|_{RBE}$ is a pure state (conditioned on $X$).

We proceed with applying the Schmidt decomposition to the pure states $\{|\phi^{x}\rangle_{RBE}\}_{x\in \mathcal{X}}$ with respect to the cut $R|BE$ . Let $\{|y^{x}\rangle_{R}\}$ and $|\psi^{x,y}\rangle_{BE}$ be orthonormal bases for $R$ and $BE$ systems. Then from Schmidt decomposition we have that
\begin{align*}
|\phi^{x}\rangle_{RBE}=\sum_{y}\sqrt{p_{Y|X}(y|x)}|y^{x}\rangle_{R}\otimes |\psi^{x,y}\rangle_{BE}.
\end{align*}
We want to get a decoherent version of the state $|\phi^{x}\rangle_{RBE}$ by measuring the $R$ system in the basis $\{|y^{x}\rangle_{R}\}$. Since after the measurement, $R$ system becomes a classical system, hereafter we show it by $Y$. Let $|\bar{\phi}^{x}\rangle_{YBE}$ denote the decoherent state resulting from the measurement, then
\begin{align*}
\bar{\phi}^{x}_{YBE}=\sum_{y}p_{Y|X}(y|x)|y^{x}\rangle\langle y^{x}|_{Y}\otimes |\psi^{x,y}\rangle\langle \psi^{x,y}|_{BE},
\end{align*}
and let the decoherent state $\bar{\sigma}_{XRBE}$ be as follow:
\begin{align*}
\bar{\sigma}_{XYBE}=\sum_{x}&p_{X}(x)|x\rangle\langle x|_{X}\\
&\otimes\sum_{y}p_{Y|X}(y|x)|y^{x}\rangle\langle y^{x}|_{Y}\otimes |\psi^{x,y}\rangle\langle \psi^{x,y}|_{BE}.
\end{align*}
This state is the same as was held by Bob and Eve after decoding for the public message. If the correctness of the following equality can be proven, which turns out to be straightforward, we can argue about the correctness of our claim,
\begin{align}
\label{decoh}
I(R\rangle BX)_{\sigma}=I(Y;B|X)_{\bar{\sigma}}-I(Y;E|X)_{\bar{\sigma}}.
\end{align}
The right-hand side of (\ref{decoh}) can be expanded as follow:
\begin{align*}
I&(Y;B|X)_{\bar{\sigma}}-I(Y;E|X)_{\bar{\sigma}} \\
&\stackrel{(a)}{=}H(B|X)-H(B|X,Y)-H(E|X)+H(E|X,Y) \nonumber \\
&\stackrel{(b)}{=}H(B|X)-H(E|X),
\end{align*}
where $(a)$ follows by the definition of the conditional mutual information and $(b)$ is due to the fact that conditioned on $X$ and $Y$, the state on $BE$ is a pure state. Observe the last expression is a function solely of the density operator given in (\ref{icoh}). It is evident that for the regularized formula, we consider $n$-fold states in our proof instead. This proves our claim.
\section{Conclusion}
We studied the one-shot capacity of a quantum channel for simultaneous transmission of classical and quantum information. Our main tools are position-based decoding and convex-split lemma. We first consider the problem of simultaneous transmission of public and private classical information and then we discussed that the private rate can be translated into quantum capacity. We also provided converse bounds. By evaluating our achievability and converse bounds in asymptotic i.i.d. regime, we recovered the well-known results in the literature.
\appendices
\section{Derandomization of the code}
We aim to derandomize the assisted code. As mentioned in the introductory section, this development follows the procedure used in \cite{Wil17} and \cite{HQM2017}. We start with the public message. We saw that the optimal operator $\Pi_{XB}$ is such that $\textrm{Tr}\{\Pi_{XB}\rho_{XB}\}\geq1-(\epsilon-\delta)$ and $\textrm{Tr}\{\Pi_{XB}(\rho_{X}\otimes\rho_{B})\}=2^{-I_{H}^{\epsilon-\delta}(X;B)_{\rho}}$, we rewrite the two error types with slightly different notations as follows :
\begin{align*}
\textrm{Tr}\{\Pi_{XB}\rho_{XB}\}&=\textrm{Tr}\left\{\Pi_{XB}\sum_{x}p_{X}(x)|x\rangle\langle x|_{X}\otimes\rho_{B}^{x}\right\}
\end{align*}
\begin{align*}
&=\sum_{x}p_{X}(x)\textrm{Tr}\{\langle x|\Pi_{XB}|x\rangle_{X}\rho_{B}^{x}\}\\
&=\sum_{x}p_{X}(x)\textrm{Tr}\{W_{B}^{x}\rho_{B}^{x}\},
\end{align*}
in which the operator $W_{B}^{x}$ is defined as $W_{B}^{x}\coloneqq\langle x|\Pi_{XB}|x\rangle_{X}$. In an analogous way, we have that 
\begin{align*}
\textrm{Tr}\{\Pi_{XB}(\rho_{B}\otimes\rho_{X})\}&=\textrm{Tr}\left\{\Pi_{XB}\sum_{x}p_{X}(x)|x\rangle\langle x|_{X}\otimes\rho_{B}\right\}\\
&=\sum_{x}p_{X}(x)\textrm{Tr}\{\langle x|\Pi_{XB}|x\rangle_{X}\rho_{B}\}\\
&=\sum_{x}p_{X}(x)\textrm{Tr}\{W_{B}^{x}\rho_{B}\}.
\end{align*}
These expressions imply that it is sufficient to take the optimal test to be $\Pi_{XB}=\sum_{x}|x\rangle\langle x|_{X}\otimes W_{B}^{x}$ with aforementioned $W_{B}^{x}$; In other words, the test $\Pi_{XB}$ can achieve the same error probability as any other $\Pi_{XB}$ would do.
We proceed with dissecting each term involved in $\textrm{Tr}\{(\mathbbm{1}_{X^{|\mathcal{M}|}B}-\Lambda_{X^{|\mathcal{M}|}B}^{m})\rho_{X^{\otimes|\mathcal{M}|}B}^{m,(l,k)}\}$ where $\rho_{X^{\otimes|\mathcal{M}|}B}^{m,(l,k)}$ is given in (\ref{rho1}).
\begin{figure*}[!t]
\begin{align}
\label{rho1}
\rho_{X^{\otimes|\mathcal{M}|}B}^{m,(l,k)}\coloneqq\rho_{X}^{1}...\otimes\rho_{XB}^{m,(l,k)}\otimes...\otimes\rho_{X}^{|\mathcal{M}|}
=\sum_{x_{1},...,x_{|\mathcal{M}|}}p_{X}(x_{1})...p_{X}(x_{|\mathcal{M}|})|x_{1}....x_{|\mathcal{M}|}\rangle\langle x_{1}...x_{|\mathcal{M}|}|_{X_{1}...X_{|\mathcal{M}|}}\otimes\rho_{B}^{m,(l,k)}.
\end{align}
\hrulefill
\end{figure*}
By assuming the particular structure for the optimal test operator that we just introduced, the operator $\Gamma^{m}_{X^{|\mathcal{M}|}B}$ appears as given in (\ref{cq2}).
\begin{figure*}[!t]
\begin{align}
\label{cq2}
\nonumber
&\Gamma^{m}_{X^{|\mathcal{M}|}B}=\mathbbm{1}_{X}^{1}\otimes...\otimes T_{XB}^{m}\otimes...\otimes \mathbbm{1}_{X}^{|\mathcal{M}|}
=\sum_{x_{1}}|x_{1}\rangle\langle x_{1}|_{X}\otimes...\otimes\left(\sum_{x_{m}}|x_{m}\rangle\langle x_{m}|_{X}\otimes W_{B}^{x_{m}}\right)\otimes...\otimes\sum_{x_{|\mathcal{M}|}}|x_{|\mathcal{M}|}\rangle\langle x_{|\mathcal{M}|}|_{X} \\
&=\sum_{x_{1}...x_{|\mathcal{M}|}}|x_{1}...x_{|\mathcal{M}|}\rangle\langle x_{1}...x_{|\mathcal{M}|}|_{X}\otimes W_{B}^{x_{m}}.
\end{align}
\hrulefill
\end{figure*}
And
\begin{align*}
&\left(\sum_{m'=1}^{|\mathcal{M}|}\Gamma_{X^{|\mathcal{M}|}B}^{m'}\right)^{-\frac{1}{2}} \\
&=\left(\sum_{m'=1}^{|\mathcal{M}|}\sum_{x_{1}...x_{|\mathcal{M}|}}|x_{1}...x_{|\mathcal{M}|}\rangle\langle x_{1}...x_{|\mathcal{M}|}|_{X}\otimes W_{B}^{x_{m'}}\right)^{-\frac{1}{2}}\\
&=\left(\sum_{x_{1}...x_{|\mathcal{M}|}}|x_{1}...x_{|\mathcal{M}|}\rangle\langle x_{1}...x_{|\mathcal{M}|}|_{X}\otimes\sum_{m'=1}^{|\mathcal{M}|}W_{B}^{x_{m'}}\right)^{-\frac{1}{2}}\\
&=\sum_{x_{1}...x_{|\mathcal{M|}}}|x_{1}...x_{|\mathcal{M}|}\rangle\langle x_{1}...x_{|\mathcal{M}|}|_{X}\otimes\left(\sum_{m'=1}^{|\mathcal{M}|}W_{B}^{x_{m'}}\right)^{-\frac{1}{2}},
\end{align*}
and finally
\begin{align*}
\Lambda_{X^{|\mathcal{M}|}B}^{m}&=\left(\sum_{m'=1}^{|\mathcal{M}|}\Gamma_{X^{|\mathcal{M}|}B}^{m}\right)^{-\frac{1}{2}}\Gamma_{X^{|\mathcal{M}|}B}^{m'}\left(\sum_{m'=1}^{|\mathcal{M}|}\Gamma_{X^{|\mathcal{M}|}B}^{m'}\right)^{-\frac{1}{2}}\\ 
&=\sum_{x_{1}...x_{|\mathcal{M}|}}|x_{1}...x_{|\mathcal{M}|}\rangle\langle x_{1}...x_{|\mathcal{M}|}|_{X}\otimes\Delta_{B}^{m},
\end{align*}
where
\begin{equation*}
\Delta_{B}^{m}\coloneqq\left(\sum_{m'=1}^{|\mathcal{M}|}W_{B}^{x_{m'}}\right)^{-\frac{1}{2}}W_{B}^{x_{m}}\left(\sum_{m'=1}^{|\mathcal{M}|}W_{B}^{x_{m'}}\right)^{-\frac{1}{2}}.
\end{equation*}
Note that the obtained POVM, $\{\Delta_{B}^{m}\}_{m=1}^{|\mathcal{M}|}$, can be completed by adding $\Delta_{B}^{0}=\mathbbm{1}-\sum_{m'=1}^{|\mathcal{M}|}\Delta_{B}^{m'}$.
By putting everything that has derived so far into the error term, we will have:
\begin{align*}
&\textrm{Tr}\{(\mathbbm{1}_{X^{|\mathcal{M}|}B}-\Lambda_{X^{|\mathcal{M}|}B}^{m})\rho_{X^{\otimes|\mathcal{M}|}B}^{m,(l,k)}\}\\
&\hspace*{1cm}=\sum_{x_{1},...,x_{|\mathcal{M}|}}p_{X}(x_{1})...p_{x_{|\mathcal{M}|}}\textrm{Tr}\{(\mathbbm{1}_{B}-\Delta_{B}^{m})\rho_{B}^{x_{m},(l,k)}\}.
\end{align*}
By assuming a uniform distribution on the message set, averaging it over all messages results in
\begin{align*}
&\frac{1}{|\mathcal{M}|}\sum_{m=1}^{|\mathcal{M}|}\textrm{Tr}\{(\mathbbm{1}_{X^{|\mathcal{M}|}B}-\Lambda_{X^{|\mathcal{M}|}B}^{m})\rho_{X^{\otimes|\mathcal{M}|}B}^{m,(l,k)}\}\\
&=\frac{1}{|\mathcal{M}|}\sum_{m=1}^{|\mathcal{M}|}\sum_{x_{1},...,x_{|\mathcal{M}|}}p_{X}(x_{1})...p_{X}(x_{|\mathcal{M}|})\\
&\hspace*{4.5cm}\times\textrm{Tr}\{(\mathbbm{1}_{B}-\Delta_{B}^{m})\rho_{B}^{x_{m},(l,k)}\}\\
&=\sum_{x_{1},...,x_{|\mathcal{M}|}}p_{X}(x_{1})...p_{X}(x_{|\mathcal{M}|})\\
&\hspace*{3.5cm}\times\left[\frac{1}{|\mathcal{M}|}\textrm{Tr}\{(I_{B}-\Delta_{B}^{m})\rho_{B}^{x_{m},(l,k)}\}\right],
\end{align*}
the last expression above shows averaging over all codebooks and we know that
\begin{align*}
&\sum_{x_{1},...,x_{|\mathcal{M}|}}p_{X}(x_{1})...p_{X}(x_{|\mathcal{M}|})\\
&\hspace*{3cm}\times\left[\frac{1}{|\mathcal{M}|}\textrm{Tr}\{(\mathbbm{1}_{B}-\Delta_{B}^{m})\rho_{B}^{x_{m},(l,k)}\}\right]\leq\epsilon,
\end{align*}
which in turn, says that there exists at least one particular set of values of $\{x_{1},...x_{|\mathcal{M}|}\}$ such that 
\begin{align}
\frac{1}{|\mathcal{M}|}\sum_{m=1}^{|\mathcal{M}|}\textrm{Tr}\{(I_{B}-\Delta_{B}^{m})\rho_{B}^{x_{m},(l,k)}\}\leq\epsilon.
\end{align}  
This conclusion is known as the \textit{Shannon trick}. The sequence $\{x_{1}...x_{|\mathcal{M}|}\}$ serves as the codebook used to transmit the public message. 

As for the second part, we take (\ref{p-Eve1}) and average over all private messages as given in (\ref{pricon}).
\begin{figure*}[!t]
\begin{align}
\label{pricon}
\nonumber
\epsilon+\sqrt{\epsilon'}\geq\frac{1}{|\mathcal{L}|}\sum_{\ell=1}^{|\mathcal{L}|}\sum_{x,y_{11},...,y_{|\mathcal{L}||\mathcal{K}|}}p_{X}(x)p_{Y|X}(y_{11}|x)...p_{Y|X}(y_{|\mathcal{L}||\mathcal{K}|}|x)
\left[\frac{1}{2}\left\|\mathcal{D'}_{B\rightarrow\hat{L}}^{2}\left(\frac{1}{|\mathcal{K}|}\sum_{k=1}^{|\mathcal{K}|}\sigma_{BE}^{x,m,y_{\ell k}}\right)-|\ell\rangle\langle\ell|_{\hat{L}}\otimes\tilde{\sigma}_{E}^{x,m}\right\|_{1}\right]\\
=\sum_{x,y_{11},...,y_{|\mathcal{L}||\mathcal{K}|}}p_{X}(x)p_{Y|X}(y_{11}|x)...p_{Y|X}(y_{|\mathcal{L}||\mathcal{K}|}|x)
\left(\frac{1}{|\mathcal{L}|}\sum_{l=1}^{|\mathcal{L}|}\left[\frac{1}{2}\left\|\mathcal{D'}_{B\rightarrow\hat{L}}^{2}\left(\frac{1}{|\mathcal{K}|}\sum_{k=1}^{|\mathcal{K}|}\sigma_{BE}^{x,m,y_{\ell,k}}\right)-|\ell\rangle\langle\ell|_{\hat{L}}\otimes\tilde{\sigma}_{E}^{x,m}\right\|_{1}\right]\right).
\end{align}
\hrulefill
\end{figure*}
And we again employ Shannon trick to conclude that there exists at least one sequence of values $(y_{1,1}...y_{|\mathcal{L}|,|\mathcal{K}|}|x)$ such that equation (\ref{shntrick}) holds.
\begin{figure*}[!t]
\begin{equation}
\label{shntrick}
\frac{1}{|\mathcal{L}|}\sum_{l=1}^{|\mathcal{L}|}\left[\frac{1}{2}\left\|\mathcal{D'}_{B\rightarrow\hat{L}}^{2}\left(\frac{1}{|\mathcal{K}|}\sum_{k=1}^{|\mathcal{K}|}\sigma_{BE}^{x,m,y_{\ell,k}}\right)-|\ell\rangle\langle \ell|_{\hat{L}}\otimes\tilde{\sigma}_{E}^{x,m}\right\|_{1}\right]\leq\epsilon+\sqrt{\epsilon'}.
\end{equation}
\hrulefill
\end{figure*}

We can now argue that there exist values $(x_{1}...x_{|\mathcal{M}|})$ serving as \textit{public codebook} for the transmission of the public message and conditioned on a particular codeword of the public codebook, there exist values $(y_{1,1}...y_{|\mathcal{L}|,|\mathcal{K}|})$ serving as \textit{private codebook} ensuring that the privacy criterion holds. Now we have a codebook of size $|\mathcal{M}||\mathcal{L}||\mathcal{K}|$, $\{x_{1},...,x_{|\mathcal{M}|},y_{1},...,y_{|\mathcal{L}||\mathcal{K}|}\}$, that is publicly available serving as our deterministic codebook for simultaneous transmission of public and private messages.
\section{One-shot Quantum Capacity: Imitating Devetak's Asymptotic Proof }
As we mentioned in the introduction, a one-shot version of Devetak's asymptotic proof of quantum capacity follows along the same lines \cite{Dev05} . Here we briefly outline the proof and the general idea. We shall freely use the notation introduced so far. We have now seen that there exists a good codebook $\{x(\ell,k)\}_{\ell\in\mathcal{L},k\in\mathcal{K}}$ selected from a distribution $p(x)$ and a corresponding POVM $\{\Omega_{B}^{\ell,k}\}$ such that once Alice transmits a state corresponding to $x(\ell,k)$ over the channel $\mathcal{N}_{A\rightarrow BE}$, Bob is able to reliably work out both Alice's message $\ell$ and the local key $k$ and at the same time, Eve happens to learned very little about Alice's message $\ell$.
This holds true for the rates of the private message $\log_{2}|\mathcal{L}|$ and the local randomness $\log_{2}|\mathcal{L}|$ as are specified. Now a quantum code can be obtained by a \enquote{making coherent} of this code (see \cite{DHW} for coherifying general protocols).

 The first idea of making protocols coherent is that classical
words/letters $x$ become basis states $\ket{x}$ of the Hilbert space. Functions $f:x\rightarrow f(x)$ thus induce linear operators on Hilbert space, but only permutations (one-to-one functions) are really interesting, since they give rise to unitaries (isometries, resp.). The second idea is thus to make classical computations first reversible, by extending them to one-to-one functions. The last step is to use the local decodings, which exist by the classical theorem. In summary, "making coherent" means we can take a classical
protocol working on letters and turn it into a bunch of
unitaries acting as permutations on the basis states, and
that we can run perfectly well on superpositions.

From the recipe outlined above, Alice's messages $\ell\in\mathcal{L}$ become a basis $\{\ket{\ell}_{A_{1}}\}_{\ell\in\mathcal{L}}$ of the Hilbert space. Suppose that Alice shares a state $\ket{\varphi}_{RA_{1}}$ with a reference system $R$:
\begin{align*}
\ket{\varphi}_{RA_{1}}\coloneqq\sum_{i,\ell\in\mathcal{L}}\alpha_{i,\ell}\ket{i}_{R}\ket{\ell}_{A_{1}},
\end{align*}
where $\ket{i}_{R}$ and $\ket{\ell}_{A_{1}}$ are some orthonormal bases for $R$ and $A_{1}$, respectively. A number of different information-processing tasks can be considered as quantum communications. The strongest definition of quantum capacity, however, corresponds to a task known as \textit{entanglement transmission}. According to this task, Alice aims to transfer her share of entanglement with a reference system to Bob with Alice no longer entangled with the reference, i.e., the following state:
\begin{align*}
\ket{\varphi}_{RB}\coloneqq\sum_{i,\ell\in\mathcal{L}}\alpha_{i,\ell}\ket{i}_{R}\ket{\ell}_{B}, 
\end{align*}   
where now Bob holds the $B$ system. Suppose there is an ensemble of quantum states $\{p(x),\ket{\psi^{x}}_{A}\}$. Alice uses her classical code to create a quantum codebook whose codewords are as follows:
\begin{align*}
\ket{\phi^{\ell}}_{A}\coloneqq\frac{1}{\sqrt{|\mathcal{K}|}}\sum_{k\in\mathcal{K}}e^{\gamma_{\ell,k}}\ket{\psi^{x(\ell,k)}}_{A},
\end{align*}
where the states $\ket{\psi^{x(\ell,k)}}_{A}$ are from the aforementioned ensemble and $x(\ell,k)$ belong to the (classical) private codebook. Alice's action would be to coherently copy the value of $\ell$ in register $A_{1}$ to another register $A_{2}$. She then applies some isometric encoding from $A_{2}$ register to $A$. These two steps are performed with the following map:
\begin{align*}
\left(\sum_{\ell}\ketbra{\ell}_{A_{2}}\otimes\ket{\phi^{\ell}}_{A}\right)\left(\sum_{\ell}\ketbra{\ell}_{A_{1}}\otimes\ket{\ell}_{A_{2}}\right),
\end{align*}
Alice then transmits the codeword over the channel giving rise to the following state:
\begin{align*}
\sum_{i,\ell\in\mathcal{L}}\alpha_{i,\ell}\ket{i}_{R}\ket{\ell}_{A_{1}}\ket{\phi^{\ell}}_{BE}.
\end{align*}
From the classical protocol we know that Bob can detect both the message $\ell$ and the local key $k$ with high probability. Bob constructs a coherent version of his POVM as follows:
\begin{align*}
\sum_{\ell,k}\sqrt{\Omega_{B}^{\ell,k}}\otimes\ket{\ell}_{B_{1}}\ket{k}_{B_{2}}.
\end{align*}

From gentle measurement lemma, the state after Bob's decoding will be close to the following state:
\begin{align*}
\sum_{i,\ell}\sum_{k}\frac{1}{\sqrt{|\mathcal{K}|}}\alpha_{i,\ell}\ket{i}_{R}\ket{\ell}_{A_{1}}e^{\delta_{\ell,k}}\ket{\phi^{x(\ell,k)}}_{BE}\ket{\ell}_{B_{1}}\ket{k}_{B_{2}}.
\end{align*}
On the other hand, from secrecy requirement, it can be seen that there exists some isometry on Bob's $B$ and $B_{2}$ systems such that after its application, Eve's system will be decoupled from the rest and the following state will result:
\begin{align*}
\sum_{i,\ell}\alpha_{i,\ell}\ket{i}_{R}\ket{\ell}_{A_{1}}\ket{\ell}_{B_{1}}.
\end{align*}
So far they have successfully implemented an approximate coherent channel from systems $A_{1}$ to $A_{1}B_{1}$. Alice is allowed to use a forward classical channel to communicate with Bob in order to turn the above coherent channel to a quantum channel. Alice's strategy is to first perform a Fourier transform on the register $A_{1}$ then measure the register in the computational basis and communicate the classical output to Bob. Bob will perform a controlled unitary based on the classical letter he received and the desired state will be achieved. Note than it can be shown that there exists a scheme that does not require the use of this forward classical channel.

\section*{Acknowledgment}
The first author is indebted to Andreas Winter for his constant advice; In regards to this work, the part on making protocols coherent is due to his private communications with Prof. Winter. He also sincerely thanks the centre for quantum software and information at the University of Technology Sydney for their hospitality and support while part of this work was done. The authors are grateful to M. M. Wilde for his comments on the first version of this work. FS acknowledges partial financial support 
by the Baidu-UAB collaborative project 'Learning of Quantum 
Hidden Markov Models', the Spanish MINECO (project FIS2016-86681-P) 
with the support of FEDER funds, and the Generalitat de Catalunya 
(project 2017-SGR-1127). M.-H.~Hsieh was supported by an ARC Future Fellowship under Grant FT140100574 and by US Army Research Office for Basic Scientific Research Grant W911NF-17-1-0401. The work of F.~Salek and J.~R.~Fonollosa has been funded by the \enquote{Ministerio de Economía, Industria y Competitividad} of the Spanish Government, ERDF funds (TEC2013-41315-R, TEC2015-69648-REDC, TEC2016-75067-C4-2-R) and the Catalan Government (2017 SGR 578  AGAUR). A.~Anshu and R.~Jain are supported by the Singapore Ministry of Education through the Tier 3 Grant \enquote{Random numbers from quantum processes} MOE2012-T3-1-009. R.~Jain is also supported by the VAJRA Grant, Department of Science and Technology, Government of India and by the NRF2017-NRF-ANR004 VanQuTe grant.

\ifCLASSOPTIONcaptionsoff
  \newpage
\fi

\end{document}